\documentclass[12pt,a4paper]{article}
\usepackage[T1]{fontenc}
\usepackage[utf8]{inputenc}
\usepackage{lmodern}
\usepackage[a4paper,margin=2.5cm]{geometry}
\usepackage{microtype}
\usepackage{amsmath,amssymb,amsthm,mathtools}
\usepackage[authoryear,round]{natbib}
\usepackage{xurl}
\usepackage{setspace}
\usepackage[unicode,bookmarksnumbered,hidelinks,hypertexnames=false]{hyperref}
\hypersetup{
 pdftitle={Individual incentives and uniform gains under symmetric constraints},
 pdfauthor={Siwei Chen},
 pdfsubject={Strategy-proof allocation under symmetric resource constraints},
 pdfkeywords={single-peaked preferences, strategy-proofness, uniform gains, symmetric division}
}
\newtheorem{theorem}{Theorem}

\newtheorem{corollary}{Corollary}
\newtheorem{lemma}{Lemma}
\newtheorem{claim}{Claim}
\theoremstyle{definition}

\theoremstyle{remark}

\numberwithin{equation}{section}
\DeclareMathOperator{\med}{med}
\DeclareMathOperator{\clamp}{clamp}
\DeclareMathOperator{\aff}{aff}

\newcommand{\R}{\mathbb R}

\newcommand{\one}{\mathbf 1}
\AtBeginDocument{\typeout{MANUSCRIPT-BASELINESKIP: \the\baselineskip}}
\renewenvironment{abstract}
 {\begin{center}\bfseries Abstract\end{center}\noindent\ignorespaces}
 {\par}
\title{Individual incentives and uniform gains\\under symmetric constraints}
\author{Siwei Chen\thanks{Lingnan College, Sun Yat-sen University, Guangzhou, China.
Email: \href{mailto:chensw6@mail.sysu.edu.cn}{chensw6@mail.sysu.edu.cn}.
ORCID: \href{https://orcid.org/0000-0002-8486-2251}{0000-0002-8486-2251}.}}
\date{November 9, 2026}
\begin{document}
\maketitle

\begin{abstract}
We study the division of a fixed total under symmetric joint constraints when agents have continuous, strictly single-peaked preferences over their own shares. Each agent reports a complete preference, including comparisons across the two sides of her peak. For any finite population and any nonempty closed convex permutation-invariant feasible set, we characterize the uniform-gains rule by efficiency, symmetry, individual strategy-proofness, and own-coordinate peak-onlyness. The last condition requires that a change of preference leaving an agent's peak fixed also leave her own share fixed; it places no direct restriction on others' shares. The characterization separates uniqueness from continuity and yields a continuous characterization with individual incentives for compact feasible sets that are polyhedral or relatively full-dimensional and strictly convex. We then show that, for every population of two through six agents and every nonempty compact convex permutation-invariant feasible set, efficiency, symmetry, and individual strategy-proofness alone characterize uniform gains. The proof recovers allocations from the shares attainable by unilateral reports, feasible transfers, and endpoint reports. A common induction applies at each population, with additional starting configurations for five and six agents.
\end{abstract}


\section{Introduction}\label{sec:intro}

When does truthful individual reporting determine how a resource should be divided? In the classical division problem, efficiency, anonymity, and strategy-proofness identify the uniform rule. That conclusion is particularly informative because the three requirements concern different aspects of an allocation: the use of resources, the treatment of agents, and their incentives. It also relies on a simple feasibility constraint. Once shares must satisfy joint restrictions, it is less clear whether the same requirements still select a unique rule.

We consider a fixed total and any closed convex feasible set that treats agents interchangeably. These constraints allow, for example, bounds on the dispersion of shares or on the total received by any group of a given size. Each agent cares only about her own share and has a continuous, strictly single-peaked preference. Reports contain the entire preference, so agents with the same preferred share may disagree about comparisons across that share. A mechanism can therefore respond to more information than the vector of peaks.

The allocation rule of interest is uniform gains, which agrees with the classical uniform rule on the division simplex. It begins at equal division and equalizes, in the leximin sense, agents' movements toward their preferred shares, subject to feasibility. \citet{moulin17} introduced the rule in a general one-dimensional mechanism design model. In symmetric division problems, his Proposition~2 characterizes it by efficiency, symmetry, continuity, and strong group strategy-proofness. His footnote~13 asks whether individual strategy-proofness can replace the coalitional requirement.

Joint constraints make this question substantive. In the classical problem, every efficient allocation is one-sided: either no agent receives more than her peak or no agent receives less. That property can fail under joint restrictions. Moreover, individual strategy-proofness directly controls only the share of the agent changing her report. Her deviation may redistribute resources among the other agents even when her own share is unchanged. A uniqueness proof must recover those other shares from the economic axioms.

Our first theorem isolates a sufficient restriction on the use of an agent's information. Own-coordinate peak-onlyness requires two reports with the same peak to give their sender the same share, holding the other reports fixed. For every finite population, efficiency, symmetry, individual strategy-proofness, and this restriction characterize uniform gains on every nonempty closed convex permutation-invariant subset of a budget hyperplane.

This theorem also clarifies the role of continuity. The first part of Moulin's continuity argument implies own-coordinate peak-onlyness from individual strategy-proofness and continuity alone. Combining that implication with our first theorem gives the continuous characterization in the regular compact class, with individual strategy-proofness in place of strong group strategy-proofness. Continuity thus supplies the own-allocation restriction used in the identification argument.

Our second theorem dispenses with that informational restriction for every population of two through six agents. On any nonempty compact convex permutation-invariant feasible set, efficiency, symmetry, and individual strategy-proofness alone identify uniform gains. The theorem allows flat faces and all continuous strictly single-peaked preference shapes independently across agents. In this population range, the three axioms consequently imply own-coordinate peak-onlyness and the stronger incentive properties of the identified rule.

The proof uses a common set of arguments at each fixed population. With own-coordinate peak-onlyness, an agent's allocation is a scalar function of her peak, and strategy-proofness gives a median identity. Without that restriction, we instead work with the shares actually attainable by changing one complete report while fixing all other complete reports. Endpoint reports provide attainable bounds; feasible transfers and efficiency determine the remaining coordinates. The additional arguments needed for five and six agents identify specific starting configurations for this same induction. Every deviation takes place within the original population and feasible set.

These results provide two complementary answers to the identification question. An own-allocation restriction suffices for arbitrary finite populations, and the three economic axioms themselves suffice for populations through six on compact feasible sets. Whether the three axioms suffice for larger populations remains open.

\subsection{Related literature}\label{sec:lit}

The classical division problem is the starting point. \citet{sprumont91} characterizes the uniform rule using efficiency and strategy-proofness together with anonymity or no envy. \citet{ching94} obtains a characterization using equal treatment of equals. The present paper keeps a single commodity and single-peaked preferences while allowing symmetric joint constraints beyond the classical simplex.

The closest comparison is \citet[Proposition~2]{moulin17}. His construction establishes the existence and desirable properties of uniform gains on a much larger class of closed convex sets. We retain his rule and focus on its uniqueness in symmetric fixed-total problems. The first theorem weakens the incentive and informational premises used in the uniqueness argument and separates that argument from regularity conditions ensuring continuity. The second theorem removes the informational condition on general compact feasible sets through six agents. Both theorems concern fixed-total division problems.

The scalar part of our argument is related to the median representations of \citet{moulin80} and \citet{bgs93}, and to uncompromisingness in \citet{borderjordan83}. Those results concern a common scalar outcome. Here each agent consumes one coordinate of a vector, and changes in the other coordinates must be identified separately. The distinction between individual and coalitional incentive requirements is also central to \citet{bbm10,bbm16}. Our argument uses symmetry and efficiency to recover the other coordinates after a unilateral incentive comparison identifies the reporting agent's share.

Constrained division also includes models with individual capacity restrictions, such as \citet{bergantinos15}. Our feasibility restriction is instead a symmetric joint condition on the allocation vector. The shared fixed-total geometry allows averaging and transfers between coordinates, while the incentive comparisons continue to concern one reporting agent at a time.

\section{The allocation problem}\label{sec:model}

Fix a finite set of agents $N=\{1,\ldots,n\}$, with $n\geq2$, and a total $\beta\in\R$. Let
\[
H_\beta=\left\{x\in\R^n:\sum_{i\in N}x_i=\beta\right\}.
\]
The feasible set $X$ is a nonempty closed convex subset of $H_\beta$, invariant under every permutation of coordinates. For a permutation $\sigma$ of $N$, write $x^\sigma_{\sigma(i)}=x_i$. Thus $x\in X$ implies $x^\sigma\in X$. Averaging all permutations of a feasible vector gives the feasible equal split
\[
\omega=\bar\omega\one,\qquad \bar\omega=\frac{\beta}{n}.
\]
All coordinate projections of $X$ coincide; denote this interval by $I$. Lemma~\ref{lem:dichotomy} shows that $X$ is either compact or $H_\beta$, so $I$ is a compact interval or $\R$.

For an illustration from \citet[Section~8.2]{moulin17}, four partners divide $100$ shares, with nonnegative holdings and a bound of $66$ on the combined holding of any pair. At peaks $(10,15,35,40)$, the allocation $(17,17,30,36)$ is efficient even though the first two partners receive more than their peaks and the last two receive less. A weak improvement must weakly reduce the first two shares and weakly increase the last two. The binding bound on the last pair fixes both of their shares; the budget then fixes the first two. Thus the one-sidedness of efficient allocations in the classical division problem need not survive joint constraints.

A preference $R_i$ on $I$ is complete, transitive, continuous, and strictly single-peaked. It has a unique peak $p(R_i)$, with
\[
a<b\leq p(R_i)\ \Longrightarrow\ b\,P_i\,a,
\qquad
p(R_i)\leq a<b\ \Longrightarrow\ a\,P_i\,b,
\]
where $P_i$ is its strict part. Here continuity is a property of the preference relation. Write $\mathcal S(I)$ for the full class of such preferences and $p_i=p(R_i)$. The report domain is the entire independent product $\mathcal S(I)^n$. In particular, preferences with the same peak may make different comparisons across that peak. If $I$ is a singleton, the unique allocation is understood throughout.

An allocation rule is a map $F:\mathcal S(I)^n\to X$. Feasibility is built into this definition. The rule is \emph{efficient} if, at every profile $R$, there is no $y\in X$ such that $y_i\,R_i\,F_i(R)$ for all $i$ and $y_j\,P_j\,F_j(R)$ for some $j$. It is \emph{symmetric} if
\[
F(R^\sigma)=F(R)^\sigma
\]
for every profile and permutation, where $R^\sigma_{\sigma(i)}=R_i$. Symmetry acts on the profile of complete preferences.

The rule is \emph{individually strategy-proof} if
\begin{equation}\label{eq:sp}
 F_i(R_i,R_{-i})\ R_i\ F_i(R_i',R_{-i})
 \quad\text{for every }i,\ R_i,\ R_i',\ R_{-i}.
\end{equation}
All unilateral comparisons hold the other agents' complete reports fixed. We abbreviate efficiency, symmetry, and strategy-proofness by EFF, SYM, and SP.

The rule satisfies \emph{own-coordinate peak-onlyness}, abbreviated OCPO, if
\begin{equation}\label{eq:ocpo}
 p(R_i)=p(R_i')
 \ \Longrightarrow\
 F_i(R_i,R_{-i})=F_i(R_i',R_{-i})
 \quad\text{for every }i,\ R_i,\ R_i',\ R_{-i}.
\end{equation}
This condition restricts only the reporting agent's share. Full peak-onlyness would require the entire allocation to depend only on the peak vector. Non-bossiness would require the entire allocation to stay fixed whenever a unilateral change leaves the reporting agent's share fixed.

For $a,b\in\R^n$, define their coordinate box by
\[
B(a,b)=\{x\in\R^n:\min(a_i,b_i)\leq x_i\leq\max(a_i,b_i)
\text{ for all }i\}.
\]
The \emph{uniform-gains rule} $U$ selects from $X\cap B(\omega,p)$ the allocation whose gain vector
\[
g(x)=(|x_i-\omega_i|)_{i\in N}
\]
is maximal in the leximin ordering. To compare gain vectors, sort their coordinates increasingly and use lexicographic order. This maximizes the smallest gain, then the next smallest, and so on. The gains measure movement from equal split toward each agent's peak, stopping at that peak.

The feasible gain set is nonempty and compact. On the fixed box, the gain transformation is affine in each coordinate with a fixed sign, so its image is convex. Leximin has a unique maximizer on this compact convex set, and the allocation is recovered uniquely from its gains and the direction toward each peak. Hence $U$ is well defined and peak-only. By \citet[Theorem~1]{moulin17}, it is efficient, symmetric, and strongly group strategy-proof on closed convex feasible sets with this symmetric benchmark. Strong group strategy-proofness means that no coalition can misreport so that all its members weakly gain and at least one strictly gains. In particular, $U$ satisfies SP and OCPO. We also use its defining box property:
\begin{equation}\label{eq:box}
 \min(\bar\omega,p_i)\leq U_i(R)\leq\max(\bar\omega,p_i)
 \quad\text{for every }i.
\end{equation}
The compact-case proof below gives a direct construction and verifies the canonical properties needed for the second theorem.

\section{Characterization}\label{sec:results}

\begin{theorem}\label{thm:uniqueness}
Let $n\geq2$ and let $X\subseteq H_\beta$ be nonempty, closed, convex, and invariant under every coordinate permutation. A rule on $\mathcal S(I)^n$ is efficient, symmetric, individually strategy-proof, and own-coordinate peak-only if and only if it is the uniform-gains rule $U$.
\end{theorem}

The theorem applies to arbitrary finite populations and includes the unbounded budget hyperplane. OCPO is imposed only on the rule being characterized; the conclusion identifies its entire allocation. Thus a restriction on each reporting agent's own share, together with the three economic axioms, determines all cross-agent effects of reports.

For the continuous characterization, equip the preference domain with closed convergence of preference graphs and its product topology, as in \citet[Section~5]{moulin17}. On a compact share interval, closed convergence can be described by Hausdorff convergence of the closed graphs in $I\times I$. A rule is continuous when convergence of complete preference profiles in this topology implies convergence of allocation vectors.

\begin{corollary}\label{cor:moulin}
Let $X\subseteq H_\beta$ be nonempty, compact, convex, and invariant under every coordinate permutation. Suppose $X$ is either a polytope or strictly convex of affine dimension $n-1$, with strict convexity understood relative to $H_\beta$. Then a rule on $\mathcal S(I)^n$ is efficient, symmetric, individually strategy-proof, and continuous if and only if it is $U$.
\end{corollary}

\begin{proof}
The first part of the proof of \citet[Lemma~1]{moulin17} uses only SP and continuity to obtain OCPO; footnote~9 in that paper explicitly records this implication. Theorem~\ref{thm:uniqueness} therefore gives $F=U$. Conversely, $U$ has the three economic properties, and its continuity in the stated regularity environment follows from the continuity argument for \citet[Proposition~2]{moulin17}.
\end{proof}

The corollary replaces the strong group incentive premise in Moulin's continuous characterization by individual strategy-proofness in this compact regular class. Theorem~\ref{thm:uniqueness} shows why the continuity assumption is useful and why it is unnecessary once OCPO is available.

\begin{theorem}\label{thm:low-dimensional}
Let $2\leq n\leq6$ and let $X\subseteq H_\beta$ be nonempty, compact, convex, and invariant under every coordinate permutation. A rule on the full independent domain $\mathcal S(I)^n$ is efficient, symmetric, and individually strategy-proof if and only if it is $U$.
\end{theorem}

Theorem~\ref{thm:low-dimensional} allows all complete reports independently, including distinct preferences with the same peak, and includes singleton feasible sets.

Both characterizations also imply the stronger properties of uniform gains, including strong group strategy-proofness.

\section{How individual incentives identify the allocation}\label{sec:mechanism}

The two proofs share the same economic division of labor. Strategy-proofness controls the share of a reporting agent. Symmetry makes identical endpoint reporters receive identical shares. The budget and efficiency then recover the coordinates left undetermined by the unilateral comparison. The difference is how the reporting agent's own share is identified.

\subsection{An own-allocation identity}

The following statements concern the reporting agent's own allocation, holding the other complete reports fixed.

For two real numbers $a$ and $b$, let $[a,b]$ denote the closed interval with these endpoints,
regardless of their order. The richness of the full single-peaked domain first turns an ordinal
comparison into an interval restriction.

\begin{claim}\label{lem:unanimous}
Let $I$ be a closed interval, possibly unbounded, and let $\mathcal D(p)$ be the set of all continuous
strictly single-peaked preferences on $I$ with peak $p$. For $x,y,p\in I$,
\begin{equation}
x\;R\;y\quad\text{for every }R\in\mathcal D(p)
\quad\Longleftrightarrow\quad
x\in[p,y].
\label{eq:unanimous}
\end{equation}
If $x\ne y$ and $x\in[p,y]$, every preference in $\mathcal D(p)$ strictly ranks $x$ above $y$.
\end{claim}

\begin{proof}
If $x\in[p,y]$, then $x$ lies weakly between the peak and $y$. Strict single-peakedness therefore
ranks $x$ weakly above $y$, and strictly above it when the two points differ. Conversely, suppose
$x\notin[p,y]$. If $x$ and $y$ are on the same side of $p$, strict monotonicity on that side ranks
$y$ above $x$. If they are on opposite sides, say $x<p<y$, consider the utility representation
\[
u(z)=
\begin{cases}
-a(p-z),&z\leq p,\\
-b(z-p),&z\geq p,
\end{cases}
\]
where $a,b>0$. It is continuous and strictly single-peaked at $p$, so it represents a preference in
$\mathcal D(p)$. Choosing
$b(y-p)<a(p-x)$ makes $u(y)>u(x)$. The other cross-peak ordering is symmetric. Hence unanimous weak
preference for $x$ is possible only when $x\in[p,y]$.
\end{proof}

Fix an agent $i$ and the other reports $R_{-i}$. By OCPO, the own allocation induced by a report with
peak $q$ is well defined independently of the shape of that report. Write
\[
g(q)=F_i(R_i^q,R_{-i}),
\]
where $R_i^q$ is any preference with peak $q$. The agent $i$ and the reports $R_{-i}$ are fixed
throughout and are suppressed from the notation; every statement about $g$ below is understood to
hold for each such pair. For arbitrary peaks $p,q$, put $x=g(p)$ and
$y=g(q)$. Strategy-proofness must hold for every preference with the relevant peak. Applying
Claim~\ref{lem:unanimous} in both directions gives
\begin{equation}
x\in[p,y]
\qquad\text{and}\qquad
y\in[q,x].
\label{eq:two-intervals}
\end{equation}

\begin{claim}\label{lem:monotonicity}
The function $g$ is weakly increasing; explicitly,
\[
p<q\quad\Longrightarrow\quad g(p)\leq g(q).
\]
\end{claim}

\begin{proof}
Let $p<q$ and suppose, contrary to the claim, that $x=g(p)>g(q)=y$. The first inclusion in
\eqref{eq:two-intervals} then forces $y<x\leq p$, while the second forces $q\leq y<x$. Together they
imply $q\leq p$, a contradiction.
\end{proof}

\begin{claim}\label{lem:own-uncompromising}
If $g(p)=x<p$, then $g(q)=x$ for every $q\geq x$. Symmetrically, if $g(p)=x>p$, then $g(q)=x$ for
every $q\leq x$.
\end{claim}

\begin{proof}
Consider the first case and write $y=g(q)$, where $q\geq x$. The first inclusion in
\eqref{eq:two-intervals}, together with $x<p$, implies $y\leq x$. The second inclusion and $q\geq x$
imply $y\geq x$. Thus $y=x$. The proof of the second statement reverses all inequalities.
\end{proof}

\begin{lemma}\label{lem:median}
For any $\ell\leq p\leq h$,
\begin{equation}
g(p)=\med\{g(\ell),p,g(h)\}.
\label{eq:median}
\end{equation}
\end{lemma}

\begin{proof}
If $g(p)<p$, Claim~\ref{lem:own-uncompromising} gives $g(h)=g(p)$, while
Claim~\ref{lem:monotonicity} gives $g(\ell)\leq g(p)$. Thus
$g(\ell)\leq g(h)=g(p)<p$, and the median in \eqref{eq:median} is $g(p)$. If $g(p)>p$, the same two
lemmas give $g(\ell)=g(p)\leq g(h)$, so the median is again $g(p)$. Finally, if $g(p)=p$,
monotonicity gives $g(\ell)\leq p\leq g(h)$.
\end{proof}

The median identity identifies the reporting agent's share from two endpoint allocations and her current peak. We now recover the remaining coordinates.

\subsection{Recovering the allocation on a compact feasible set}\label{subsec:compact-proof}

When $X$ is compact, the endpoints of $I$ have unique ordinal preferences: the low-endpoint preference is strictly decreasing, and the high-endpoint preference is strictly increasing. Starting from profiles consisting entirely of these endpoint preferences, we introduce interior reports one at a time. Applying~\eqref{eq:median} to each interior reporter identifies all of their shares. Symmetry, the budget, and efficiency then determine the remaining endpoint coordinates.

The following argument proves the compact branch of Theorem~\ref{thm:uniqueness}. Appendix~\ref{sec:proof} proves that every unbounded feasible set in the model is the whole budget hyperplane and completes that case.

\label{subsec:compact}
\begin{proof}[Proof of Theorem~\ref{thm:uniqueness} for compact $X$]
The common coordinate projection is a compact interval, say
$X_i=[\underline x,\overline x]$. If its endpoints coincide, then $X=\{\omega\}$ and the conclusion
is immediate. Assume $\underline x<\overline x$. A preference with peak $\underline x$ is strictly
decreasing throughout $X_i$ and is therefore ordinally unique; denote it by $L$. Similarly, denote
the unique strictly increasing preference with peak $\overline x$ by $H$. Call every report whose
peak lies in $(\underline x,\overline x)$ an interior report.

We induct on the number of interior reports in a profile. First consider a profile containing only
$L$ and $H$. If all agents report the same endpoint preference, symmetry and the budget equation give
$F(R)=U(R)=\omega$. Suppose instead that both endpoint types occur. Symmetry gives common shares
$a,b$ under $F$ and $c,d$ under $U$ for the $H$- and $L$-agents, respectively. If there are $k$
$H$-agents, feasibility yields
\[
ka+(n-k)b=kc+(n-k)d=\beta.
\]
If $a>c$, then $b<d$. Every $H$-agent strictly prefers $a$ to $c$, and every $L$-agent strictly
prefers $b$ to $d$, so $F(R)$ Pareto dominates $U(R)$, contradicting the efficiency of $U$. If
$a<c$, the reverse dominance contradicts the efficiency of $F$. Hence $a=c$ and $b=d$.

For the induction step, suppose all profiles with at most $m-1$ interior reports have been identified,
and let $R$ have $m$ interior reports. Fix an interior agent $i$, with peak $p_i$, and let $R^{i,L}$
and $R^{i,H}$ replace only her report by $L$ and $H$, respectively. The induction hypothesis gives
\[
F(R^{i,L})=U(R^{i,L})
\quad\text{and}\quad
F(R^{i,H})=U(R^{i,H}).
\]
Applying Lemma~\ref{lem:median} first to $F$ and then to $U$ gives
\begin{equation}
F_i(R)=\med\{F_i(R^{i,L}),p_i,F_i(R^{i,H})\}
=\med\{U_i(R^{i,L}),p_i,U_i(R^{i,H})\}=U_i(R).
\label{eq:compact-interior}
\end{equation}
Thus $F$ and $U$ agree coordinate by coordinate for every interior agent.

It remains to recover the endpoint coordinates. If there are no endpoint reporters, this has already
been done. If only one endpoint type is present, symmetry gives a common share for that type under
each rule, and the budget equation together with \eqref{eq:compact-interior} makes the two common
shares equal. If both types are present, write their common shares again as $a,b$ under $F$ and
$c,d$ under $U$. Since the interior coordinates agree, subtracting the two budget equations gives
\[
|N_H|(a-c)+|N_L|(b-d)=0.
\]
If $a>c$, then $b<d$; all endpoint reporters strictly prefer their allocation under $F$, while every
interior agent receives the same share under the two rules. This contradicts the efficiency of $U$.
The case $a<c$ contradicts the efficiency of $F$. Therefore the endpoint shares agree as well, which
completes the induction.
\end{proof}

\section{Characterization without peak-onlyness}\label{sec:without-ocpo}

We prove Theorem~\ref{thm:low-dimensional} by identifying endpoint configurations and then introducing interior reports at the same fixed population. Throughout this section, $X$ is compact and $2\leq n\leq6$. Let $L$ and $H$ denote the unique preferences with peaks at the lower and upper endpoints of the common projection.

\subsection{Attainable shares}\label{subsec:menus-outline}\label{u26:II-1}\label{u26:ii:menus}

Without OCPO, the function $g$ in~\eqref{eq:median} need not be well defined. The second proof therefore starts with an agent's actual menu
\[
\mathcal A_i(R_{-i})
 =\{F_i(Q_i,R_{-i}):Q_i\in\mathcal S(I)\}.
\]
The opponents are complete reports and remain fixed in this definition. Strategy-proofness says that the truthful outcome maximizes the agent's true preference over this menu. If a legal peak itself belongs to the menu, a report with that peak must receive it. More generally, a missing interval in the menu restricts how reports on the two sides can be assigned. These implications retain the information in the complete report.

\begin{lemma}
\label{u26:actual-menus}
For fixed \(n-1\) complete opponents, the menu of attainable shares is compact. Let
\(\lambda,\eta\) be its minimum and maximum, attained by \(L,H\). If an
actual pair of \(L/H\) returns equals the pair $U_i(L,R_{-i}),U_i(H,R_{-i})$, strategy-proofness pins \(F_i=U_i\) whenever \(p_i\) lies outside the open bracket
\((\lambda,\eta)\).
\end{lemma}

\begin{proof}
If attained quantities tend to \(z\), a complete report with peak \(z\) must
receive \(z\): every different return is beaten by a sufficiently close
attained quantity, by continuity and strict single-peakedness, contrary to
strategy-proofness. Thus the menu is closed in the compact projection. Since $L$ is strictly decreasing, strategy-proofness requires its truthful allocation to be the menu minimum; since $H$ is strictly increasing, its truthful allocation is the menu maximum.
If \(\lambda<p<\eta\) and \(p\) is absent, adjacent attained offers
\(l'<p<u'\) exist; finite positive piecewise-linear slopes give a full
peak-\(p\) report strictly selecting either side. At a fixed peak, two
different selected values on the same wing violate strategy-proofness, so
each selected wing has at most one value.

Lemma~\ref{lem:median} applies to the peak-only rule $U$ and gives its canonical median identity and endpoint extrema. An actual endpoint pair equal to that canonical pair
therefore supplies the stated pinning, including ties and collapsed intervals.
When $\lambda<p_i<\eta$, \(U_i=p_i\), and the menu contains offers on both strict
sides of the peak.
\end{proof}

\subsection{Endpoint configurations and the common induction}\label{subsec:common-induction}\label{u26:V-1}\label{u26:v:ladder}

Translate the total to zero and write the compact common projection as $[\ell,b]$. Let $L$ and $H$ be its unique low- and high-endpoint preferences. For nonnegative integers $m,l,h$ with $m+l+h=n$, use
\[
S_n(m,l,h)
\]
to denote the assertion that $F=U$ at every profile with $m$ interior reports, $l$ copies of $L$, and $h$ copies of $H$, in every placement of physical agents. Each interior report varies independently over its full admissible preference class. The assertion is uniform over all admissible feasible sets and rules at this population. The corresponding high-endpoint arguments reflect allocations and peaks about equal split and transform the feasible set and all preferences at the same time.

The induction preserves the population throughout. For $m,h\geq1$, Lemma~\ref{u26:singleton-induction} gives the implication
\begin{equation}\label{eq:bridge-outline}
\begin{gathered}
S_n(m-1,2,h),\qquad S_n(m-1,1,h+1),\qquad S_n(m,0,h+1)\\
\Longrightarrow\quad S_n(m,1,h),
\qquad m+1+h=n.
\end{gathered}
\end{equation}
The proof uses the identified profiles to supply actual endpoint allocations in menus with the required fixed opponents. Symmetry and efficiency then identify the entire target allocation.

For a concrete step, take $n=3$. The first two premises in~\eqref{eq:bridge-outline} are the endpoint profiles $S_3(0,2,1)$ and $S_3(0,1,2)$. The third is $S_3(1,0,2)$, obtained from those endpoint anchors by Lemma~\ref{u26:local-oe-osa}. The implication identifies $S_3(1,1,1)$. At a profile in $S_3(2,0,1)$, either interior agent can report $L$ to reach the newly identified family, or $H$ to reach $S_3(1,0,2)$. The same completion lemma therefore identifies $S_3(2,0,1)$. The general step repeats these unilateral comparisons with more interior reporters.

\begin{samepage}
\begin{lemma}
\label{u26:source-local-ladder}
Write \(S_n(m,l,h)\), or \(S(m,l,h)\) when \(n\) is fixed, and abbreviate
\[
 H_j=S_n(j,0,n-j)\quad(0\leq j\leq n-1),\qquad
 B_j=S_n(j,1,n-j-1)\quad(0\leq j\leq n-2),
\]
\[
 A_j=S_n(j,2,n-j-2)\quad(0\leq j\leq n-3).
\]
For every \(1\leq j\leq n-2\),
\begin{equation}
 H_j,\ B_{j-1},\ A_{j-1}\quad\Longrightarrow\quad B_j,\ H_{j+1}.
 \tag{LADDER}\label{u26:LADDER}
\end{equation}
\end{lemma}
\end{samepage}

\begin{proof}
The \(A\)-range is empty at \(n=2\). The subscript \(j\) counts interior reports. Apply Lemma~\ref{u26:singleton-induction} with \(m=j\) and
\(h=n-j-1\geq1\). Its three inputs are exactly the left side of
\eqref{u26:LADDER}. That lemma identifies \(B_j\) from precisely these three previously identified families. Each interior's \(L\)-target in \(H_{j+1}\) is
\(B_j\), and its \(H\)-target is \(H_j\), at the exact retained tuple; Lemma~\ref{u26:local-oe-osa} gives \(H_{j+1}\).

Since \(S_n(m,l,h)\) holds uniformly over admissible feasible sets and rules,
reflection of the entire environment gives \(S_n(m,h,l)\). It preserves \(n\) and conjugates every complete report. Endpoint
identification supplies every valid index-zero family. Lemma~\ref{u26:local-oe-osa}, using endpoint
\(L/H\)-targets, gives \(H_1\) for every \(n\geq2\). At \(n=2\) the ladder has
no steps. For \(n\geq3\), the last step is \(j=n-2\). The required \(A\)-families
are established below at each fixed population.
\end{proof}

The common induction closes the cases of two, three, and four agents directly. For five and six agents, a further argument identifies
\[
S_{h+3}(1,2,h),\qquad h\in\{2,3\}.
\]
There is one arbitrary interior report, two low-endpoint reports, and either two or three high-endpoint reports. The same parameterized proof covers both cases. For six agents, the induction also requires the configuration
\[
S_6(2,2,2).
\]
Its two arbitrary interior reports can differ even when their peaks coincide. The argument identifies their actual menus using feasible averaging, endpoint deviations, and comparisons with the supporting geometry of $X$. Once this configuration is established, the common induction closes all six-agent profiles.

The common induction reduces the five- and six-agent proofs to these specified configurations. Appendix~\ref{app:common} develops the canonical allocation, feasible-transfer arguments, and the identification lemmas used in the induction. Appendix~\ref{app:endpoints} proves the one-interior configuration for five and six agents. Appendix~\ref{app:balanced} identifies the balanced six-agent configuration, including boundary cases and equal central peaks. The next subsection combines these results to prove the theorem at each population.

\subsection{Completing the proof at each population}
\label{u26:XIV}\label{u26:xiv}

We now prove Theorem~\ref{thm:low-dimensional} using the preceding induction. All identification assertions below retain the universal quantifiers of
\S\ref{u26:v:ladder} over legal interior peaks, independent complete report
shapes, bodies, rules, and physical placements at the same fixed \(n\).
Reflection of a family \(S_n(m,l,h)\) gives the family \(S_n(m,h,l)\), obtained by applying the assertion to the entire
conjugate environment and returning to the original one.

\subsubsection{Endpoint profiles, induction, and interior profiles}
\label{u26:xiv:ladder}

Section~\ref{u26:II-3} identifies every \(m=0\) endpoint profile, including
all-\(L\) and all-\(H\). Section~\ref{u26:III-1}, with these identified endpoint targets, gives \(H_1\). For
\(1\leq j\leq n-2\), the conditional ladder already proved is
\begin{equation}
 H_j,\;B_{j-1},\;A_{j-1}
 \quad\Longrightarrow\quad
 B_j,\;H_{j+1}. \label{u26:xiv:step}
\end{equation}
Once
\[
 H_{n-1}=S_n(n-1,0,1)
 \quad\text{and its reflection}\quad
 S_n(n-1,1,0)
\]
are known, fix a profile with \(n\) interior reports. For each agent, its
actual \(H\) and \(L\) targets are these two known families at the exact
retained complete opponents. Section~\ref{u26:II-1} pins every agent whose peak
is not in the open interval between the two canonical endpoint returns. The
canonical median of Section~\ref{u26:I-4} fulfills each remaining agent at its
peak. If the actual vector differs from \(U\), feasible \(U\) leaves the
pinned agents unchanged and gives every remaining agent its unique best
quantity, with strict improvement for someone. This contradicts
\(\mathsf{EFF}\). Hence \(S_n(n,0,0)\) follows.

\subsubsection{Two, three, and four agents}\label{u26:xiv:two-four}

For \(n=2\), there is no ladder step. The endpoint profiles, \(H_1\), and its
reflection give every profile with at least one endpoint.
\S\ref{u26:xiv:ladder} gives the two-interior profile. The \(m=0,1,2\)
counts are \(3,2,1\), respectively.

For \(n=3\), the endpoint families \(A_0,B_0\) and \(H_1\) give the premises for \(j=1\), giving
\[
 B_1=S_3(1,1,1),
 \qquad
 H_2=S_3(2,0,1).
\]
Together with their reflections, the identified families at each interior count are
\[
\begin{array}{c|l}
m&\text{identified families}\\ \hline
0&\text{Base for all four endpoint pairs}\\
1&H_1,\ B_1,\text{ and the reflection of }H_1\\
2&H_2\text{ and its reflection}\\
3&\S\ref{u26:xiv:ladder}.
\end{array}
\]
Thus all \(4+3+2+1=10\) signatures are covered.

For \(n=4\), the \(j=1\) step gives
\[
 B_1=S_4(1,1,2),
 \qquad
 H_2=S_4(2,0,2).
\]
Reflection of this proved \(B_1\) is \(A_1=S_4(1,2,1)\). Consequently
\(H_2,B_1,A_1\) give the premises for \(j=2\), giving
\[
 B_2=S_4(2,1,1),
 \qquad
 H_3=S_4(3,0,1).
\]
Coverage is
\[
\begin{array}{c|l}
m&\text{identified families}\\ \hline
0&\text{Base for all five endpoint pairs}\\
1&H_1,\ B_1,\text{ and their reflections}\\
2&H_2\text{ and its reflection, and }B_2\\
3&H_3\text{ and its reflection}\\
4&\S\ref{u26:xiv:ladder}.
\end{array}
\]
These are all \(5+4+3+2+1=15\) signatures. At \(n=2\), these arguments include the empty index sets allowed in
Sections~\ref{u26:I}--\ref{u26:V}: weighted \(\mathsf{MIN}\) is vacuous,
and full-cylinder completion allows an empty low block and active set.

\subsubsection{Five agents}\label{u26:xiv:five}

For \(n=5\), the first ladder step gives \(H_2\) and \(B_1\). Section~\ref{u26:ix:BT} supplies
\[
 A_1=S_5(1,2,2)
\]
before the next induction step. With
\(H_2,B_1,A_1\), the \(j=2\) step gives
\[
 B_2=S_5(2,1,2),
 \qquad
 H_3=S_5(3,0,2).
\]
The reflection of \(B_2\) is \(A_2=S_5(2,2,1)\). Thus \(j=3\) has all its
prior premises \(H_3,B_2,A_2\), and gives
\[
 B_3=S_5(3,1,1),
 \qquad
 H_4=S_5(4,0,1).
\]

The exhaustive coverage is
\[
\begin{array}{c|l}
m&\text{identified families}\\ \hline
0&\text{Base for all six endpoint pairs}\\
1&H_1\text{ and its reflection};\ B_1\text{ and its reflection};\ A_1\\
2&H_2\text{ and its reflection};\ B_2\text{ and its reflection }(=A_2)\\
3&H_3\text{ and its reflection};\ B_3\\
4&H_4\text{ and its reflection}\\
5&\S\ref{u26:xiv:ladder}.
\end{array}
\]
These are all \(6+5+4+3+2+1=21\) signatures. This completes the five-agent proof.

\subsubsection{Six agents}\label{u26:xiv:six}

For \(n=6\), the endpoint base and first induction step give \(H_1,B_1,H_2\). Section~\ref{u26:ix} supplies \(A_1=S_6(1,2,3)\), after which the second step gives \(B_2,H_3\). These identified families, together with the endpoint replacement result in Lemma~\ref{u26:endpoint-source}, are the inputs to Appendix~\ref{app:balanced}.
Section~\ref{u26:xiii} supplies
\[
 A_2=S_6(2,2,2)
\]
from these previously established families.

Apply \S\ref{u26:v:ladder} at \(j=3\), using \(H_3,B_2,A_2\), to obtain
\[
 B_3=S_6(3,1,2),
 \qquad
 H_4=S_6(4,0,2).
\]
Reflection of \(B_3\) is \(A_3=S_6(3,2,1)\), so \(j=4\) has its actual
premises \(H_4,B_3,A_3\) and gives
\[
 B_4=S_6(4,1,1),
 \qquad
 H_5=S_6(5,0,1).
\]
The exhaustive coverage is
\[
\begin{array}{c|l}
m&\text{identified families}\\ \hline
0&\text{Base for all seven endpoint pairs}\\
1&H_1\text{ and its reflection};\ B_1\text{ and its reflection};\
  A_1\text{ and its reflection}\\
2&H_2\text{ and its reflection};\ B_2\text{ and its reflection};\ A_2\\
3&H_3\text{ and its reflection};\ B_3\text{ and its reflection }(=A_3)\\
4&H_4\text{ and its reflection};\ B_4\\
5&H_5\text{ and its reflection}\\
6&\S\ref{u26:xiv:ladder}.
\end{array}
\]
These are all \(7+6+5+4+3+2+1=28\) signatures. Every application of the singleton induction has \(h\geq1\) and uses previously established families at the same population.

\subsubsection{Translation, singleton, and converse}\label{u26:xiv:converse}

For each \(n\), the preceding cases give \(F=U\) throughout the normalized
nontrivial domain. Translate coordinates and all complete preferences back
by \(\beta/n\). Feasibility, the independent full domain, physical
symmetry, individual strategy-proofness, and whole-\(X\) efficiency are
preserved; the peak-box leximin optimizer conjugates under the same
translation. This gives uniqueness at the original total \(\beta\).

If \(X\) is a singleton, coordinate-permutation invariance and the budget
force
\[
 X=\{(\beta/n)\one\}.
\]
Every feasible rule and \(U\) take that one value, and all required
properties hold directly.

Independently, Section~\ref{u26:I-2}, Section~\ref{u26:I-4}, and Section~\ref{u26:I-5}
prove \(U\) total, feasible, symmetric, individually strategy-proof, and
efficient against all \(X\) at each population under consideration. These properties
supply the converse and prove
the theorem for every \(2\leq n\leq6\) on the stated compact class.

\section{Conclusion}\label{sec:concl}

Individual incentives can identify uniform gains under symmetric joint resource constraints. For arbitrary finite populations, own-coordinate peak-onlyness supplies a sufficient restriction on the use of reported information. For populations of two through six, the three economic axioms imply the characterization on all compact feasible sets in the model. The common proof separates the share controlled by a unilateral incentive comparison from the other shares recovered through symmetry, feasibility, and efficiency.

The remaining general question is whether the informational restriction can be removed for larger populations on arbitrary symmetric convex feasible sets.

\appendix

\section{Geometry and the unbounded case}\label{sec:proof}

This appendix completes Theorem~\ref{thm:uniqueness}. A closed convex fully symmetric fixed-total feasible set is either compact or the whole budget hyperplane. The compact proof is in Section~\ref{subsec:compact-proof}; finite artificial endpoint reports give the unbounded case below.

\subsection{An anchor transition}\label{subsec:scalar}

Fix the reporting agent and all other complete reports, and use the function $g$ defined in Section~\ref{sec:mechanism}.

We shall also need the restriction imposed by strategy-proofness when a report moves directly between
two anchor peaks.

\begin{claim}\label{lem:anchor-transition}
Fix $\ell<h$ and the other agents' reports, and set $b=g(\ell)$ and $a=g(h)$. Then either
\[
a=b,
\]
or
\begin{equation}
\ell\leq b<a\leq h.
\label{eq:anchor-transition}
\end{equation}
\end{claim}

\begin{proof}
Equation~\eqref{eq:two-intervals} gives $b\in[\ell,a]$ and $a\in[h,b]$. Suppose $a\ne b$. If
$a<b$, the first inclusion requires $a<b\leq\ell$, whereas the second requires $h\leq a<b$,
contradicting $\ell<h$. Hence $b<a$. The two inclusions then imply $\ell\leq b$ and $a\leq h$,
which proves \eqref{eq:anchor-transition}.
\end{proof}

The preceding lemmas apply to $U$ as well as to $F$: the uniform-gains rule is peak-only and hence
OCPO, and it is strategy-proof because it is strongly group strategy-proof.

\subsection{A geometric dichotomy}\label{subsec:geometry}

The two parts of the proof correspond to the only two possible geometries of a closed convex fully
symmetric fixed-sum feasible set.

\begin{lemma}\label{lem:dichotomy}
If $X\subseteq H_\beta$ is nonempty, closed, convex, and fully symmetric, then either $X$ is compact
or $X=H_\beta$.
\end{lemma}

\begin{proof}
It is enough to consider an unbounded $X$. Fix $x^0\in X$ and choose $x^m\in X$ with
$r_m=\lVert x^m-x^0\rVert\to\infty$. Along a subsequence,
\[
d^m=\frac{x^m-x^0}{r_m}\longrightarrow d,
\qquad \lVert d\rVert=1.
\]
For any $x\in X$ and $t\geq0$, convexity gives, for all sufficiently large $m$,
\[
z^m=\left(1-\frac{t}{r_m}\right)x+\frac{t}{r_m}x^m\in X.
\]
Since $z^m\to x+td$, closedness implies $x+td\in X$. Thus $d$ is a nonzero recession direction.

Let $K=\operatorname{rec}(X)$. Because $X\subseteq H_\beta$, every $v\in K$ satisfies
$\sum_i v_i=0$. Full symmetry gives $d^\sigma\in K$ for every permutation $\sigma$, and
\[
\sum_{\sigma\in S_N}d^\sigma
=(n-1)!\left(\sum_i d_i\right)\mathbf 1=0.
\]
Consequently $-d=\sum_{\sigma\ne\mathrm{id}}d^\sigma\in K$. Applying permutations also gives
$-d^\sigma\in K$, so the full permutation orbit of $d$ lies in the lineality space
$L=K\cap(-K)$.

The coordinates of $d$ are not all equal, because $d\ne0$ and $\sum_i d_i=0$. Choose $i,j$ with
$d_i\ne d_j$ and let $\tau_{ij}$ exchange these coordinates. Since $L$ is a vector space,
\[
d-d^{\tau_{ij}}=(d_i-d_j)(e_i-e_j)\in L.
\]
It follows by permutation symmetry that $e_k-e_\ell\in L$ for every $k,\ell$. These vectors span
$H_0=\{z:\sum_i z_i=0\}$, and hence $H_0\subseteq L$. For any $x\in X$ we therefore have
$x+H_0\subseteq X$. But $x+H_0=H_\beta$, so $X=H_\beta$.
\end{proof}

\subsection{The hyperplane case}\label{subsec:hyperplane}

The unbounded branch requires the one-sidedness property that is special to the unconstrained budget
hyperplane.

\begin{lemma}\label{lem:one-sided}
Let $X=H_\beta$ and let $p$ be the peak vector at a profile. Every efficient allocation $x$ satisfies
\begin{equation}
\begin{array}{lll}
\sum_i p_i>\beta&\Longrightarrow&x_i\leq p_i\quad\text{for every }i,\\
\sum_i p_i<\beta&\Longrightarrow&x_i\geq p_i\quad\text{for every }i,\\
\sum_i p_i=\beta&\Longrightarrow&x=p.
\end{array}
\label{eq:one-sided}
\end{equation}
\end{lemma}

\begin{proof}
Suppose $\sum_i p_i>\beta$ and, contrary to the first claim, $x_i>p_i$ for some $i$. Since
$\sum_k(x_k-p_k)<0$, there is a $j\ne i$ with $x_j<p_j$. Choose
$0<\varepsilon<\min\{x_i-p_i,p_j-x_j\}$ and transfer $\varepsilon$ from $i$ to $j$. The new
allocation remains in $H_\beta$ and moves both agents strictly toward their peaks, contradicting
efficiency. If $\sum_i p_i<\beta$ and $x_i<p_i$ for some $i$, then some $j\ne i$ has $x_j>p_j$;
transferring $\varepsilon$ with
$0<\varepsilon<\min\{p_i-x_i,\,x_j-p_j\}$ from $j$ to $i$ gives the symmetric contradiction. Finally,
if $\sum_i p_i=\beta$, the peak vector itself is feasible and gives every agent her unique best
share, so it Pareto dominates every different allocation.
\end{proof}

\begin{proof}[Proof of Theorem~\ref{thm:uniqueness} for $X=H_\beta$]
Fix an arbitrary target profile $R^*$, with peak vector $p$, and for $S\subseteq N$ write
$\Pi_S=\sum_{i\in S}p_i$. Choose two real numbers $\ell<h$ such that
\begin{align}
\ell&<\min\left\{\min_i p_i,
\min_{S\subsetneq N}\frac{\beta-\Pi_S}{n-|S|}\right\},
\label{eq:low-anchor}\\
h&>\max\left\{\max_i p_i,
\max_{S\subsetneq N}\frac{\beta-\Pi_S}{n-|S|}\right\}.
\label{eq:high-anchor}
\end{align}
There are only finitely many proper subsets, so such anchors exist; taking $S=\varnothing$ shows
that the two bounds already straddle $\bar\omega$. In particular,
$\ell<p_i<h$ for every $i$ and, for every proper $S$ with $r=n-|S|$,
\begin{equation}
\Pi_S+r\ell<\beta<\Pi_S+rh.
\label{eq:anchor-crossing}
\end{equation}
Choose arbitrary preferences $L$ and $H$ with peaks $\ell$ and $h$, respectively.

For $S\subseteq N$ and $A\subseteq N\setminus S$, define the mixed profile $Q(S,A)$ by
\begin{equation}
Q_i(S,A)=
\begin{cases}
R_i^*,&i\in S,\\
H,&i\in A,\\
L,&i\in N\setminus(S\cup A).
\end{cases}
\label{eq:mixed-profile}
\end{equation}
We prove by induction on $|S|$ that $F(Q(S,A))=U(Q(S,A))$ for every admissible pair $(S,A)$.

Suppose first that all profiles with fewer than $m$ target reports have been identified, and fix
$|S|=m\geq1$ and $A\subseteq N\setminus S$. For any $i\in S$, replacing $R_i^*$ by $L$ gives
$Q(S\setminus\{i\},A)$, while replacing it by $H$ gives
$Q(S\setminus\{i\},A\cup\{i\})$. The induction hypothesis identifies both profiles. Since
$\ell<p_i<h$, Lemma~\ref{lem:median}, applied to both $F$ and $U$, yields
\begin{equation}
\begin{aligned}
F_i(Q(S,A))
&=\med\{F_i(Q(S\setminus\{i\},A)),p_i,
F_i(Q(S\setminus\{i\},A\cup\{i\}))\}\\
&=\med\{U_i(Q(S\setminus\{i\},A)),p_i,
U_i(Q(S\setminus\{i\},A\cup\{i\}))\}\\
&=U_i(Q(S,A)).
\end{aligned}
\label{eq:target-coordinates}
\end{equation}
The last equality follows from SP and OCPO of $U$. Thus all coordinates
belonging to agents in $S$ have been identified. When $S=N$, this finishes the induction step. It
remains to show how to recover the anchor coordinates whenever $S\subsetneq N$. The same argument
will also supply the induction base $S=\varnothing$, for which the target-coordinate assertion is
vacuous.

Fix a proper $S$ and put $r=n-|S|$. Order the agents in $N\setminus S$ and let
$A_0\subset A_1\subset\cdots\subset A_r=N\setminus S$, with $|A_k|=k$, be the resulting nested
chain. Write $Q^k=Q(S,A_k)$. Full symmetry implies that, at a given $k$, the allocation of a target
agent depends only on $k$, all $H$-anchors receive a common share $a_k$ under $F$, and all $L$-anchors
receive a common share $b_k$. It also implies that the choice and ordering of the nested chain are
irrelevant. By \eqref{eq:target-coordinates}, with the statement interpreted vacuously when $S$ is
empty, the target coordinates under $F$ and $U$ agree at every node.

The move from $Q^{k-1}$ to $Q^k$ changes a single anchor report from $L$ to $H$. The changing agent
receives $b_{k-1}$ before the change and $a_k$ afterward. Claim~\ref{lem:anchor-transition} therefore
gives
\begin{equation}
a_k=b_{k-1}
\quad\text{or}\quad
\ell\leq b_{k-1}<a_k\leq h.
\label{eq:chain-transition}
\end{equation}
The sum of reported peaks at node $k$ is
\begin{equation}
\Pi_k=\Pi_S+kh+(r-k)\ell.
\label{eq:peak-sum-chain}
\end{equation}
It is strictly increasing in $k$, and \eqref{eq:anchor-crossing} gives
$\Pi_0<\beta<\Pi_r$.

At $k=0$, symmetry and the budget equation uniquely determine the common $L$-anchor share after the
target coordinates have been fixed. Hence $F(Q^0)=U(Q^0)$. Denote the common share by $b_0$.
The box property of $U$ gives
\begin{equation}
\ell\leq b_0\leq\bar\omega<h.
\label{eq:left-base}
\end{equation}
Now let $1\leq k\leq r$ satisfy $\Pi_k<\beta$, and suppose $Q^{k-1}$ has been identified. Its common
$L$-anchor share satisfies $b_{k-1}\leq\bar\omega<h$ by the box property. At $Q^k$,
Lemma~\ref{lem:one-sided} and the efficiency of $F$ give $a_k\geq h$. The equality alternative in
\eqref{eq:chain-transition} is therefore impossible, while the interval alternative forces
$a_k=h$. The same one-sidedness argument and the box property give $U_H(Q^k)=h$. Because
$\Pi_r>\beta$, the inequality $\Pi_k<\beta$ implies $k<r$, so at least one $L$-anchor remains; the
budget equation now identifies its common share. Thus $F(Q^k)=U(Q^k)$. Induction from the left
identifies every node with $\Pi_k<\beta$.

At the right endpoint $k=r$, symmetry and
the budget equation identify the common $H$-anchor share, so $F(Q^r)=U(Q^r)$. Call it $a_r$; the box
property gives
\begin{equation}
\ell<\bar\omega\leq a_r\leq h.
\label{eq:right-base}
\end{equation}
Let $0\leq k\leq r-1$ satisfy $\Pi_k>\beta$, and suppose $Q^{k+1}$ has been identified. Then
$a_{k+1}\geq\bar\omega>\ell$. At $Q^k$, Lemma~\ref{lem:one-sided} gives $b_k\leq\ell$.
Applying \eqref{eq:chain-transition} to the move from $Q^k$ to $Q^{k+1}$ rules out equality and forces
$b_k=\ell$. One-sidedness and the box property likewise give $U_L(Q^k)=\ell$. Since
$\Pi_0<\beta$, the inequality $\Pi_k>\beta$ implies $k>0$, so at least one $H$-anchor remains and the
budget equation identifies its common share. Backward induction identifies every node with
$\Pi_k>\beta$.

Because $\Pi_k$ is strictly increasing, at most one node satisfies $\Pi_k=\beta$. At such a node the
peak vector is feasible and gives every agent her unique best share. Efficiency alone therefore
forces both rules to select that vector. The three cases $\Pi_k<\beta$, $\Pi_k=\beta$, and
$\Pi_k>\beta$ exhaust the chain. This reasoning also covers $r=1$: the two nodes are precisely the
two budget-identified endpoints, and there are no interior chain nodes. Symmetry then extends the
conclusion from the chosen nested chain to every $A\subseteq N\setminus S$ of the same size.

For $S=\varnothing$, the chain argument supplies the induction base. For $1\leq |S|<n$, it completes
the target-coordinate step in \eqref{eq:target-coordinates} by recovering all anchor coordinates. For
$S=N$, \eqref{eq:target-coordinates} already identifies every coordinate. The induction therefore
gives $F(Q(N,\varnothing))=U(Q(N,\varnothing))$. Since this profile is $R^*$ and $R^*$ was arbitrary,
$F=U$ on $H_\beta$.
\end{proof}

\section{Canonical allocations and common lemmas}\label{app:common}

\subsection{Model, feasible transfers, and the canonical rule}
\label{u26:I}

Fix \(n\in\{2,3,4,5,6\}\) physical agents throughout every unilateral
argument. Let \(X\) be a nonempty compact convex
coordinate-permutation-invariant subset of \(H_\beta\). Each report
is a complete continuous strictly single-peaked preference on the common
projection. The rule \(F\) is total on the full Cartesian product and
satisfies feasibility, Pareto efficiency against all of \(X\), physical
permutation symmetry, and individual strategy-proofness. Every unilateral
comparison holds the other \(n-1\) complete reports fixed.

Translate by \(\beta/n\) to \(H_0\), translating every report by the
same bijection. The domain and all four properties are preserved, and \(U\)
is conjugate under this translation. If \(X\) is a singleton, symmetry makes
it \(\{0\}\). Otherwise a nonzero permutation orbit has mean zero and
unequal-coordinate swap differences spanning \(H_0\). Hence
\(\aff X=H_0\), \(0\) is relatively interior, and the common
projection is \([\ell,\bar p]\), where \(\ell<0<\bar p\). Work in this
nontrivial normalized case until the end. In allocation vectors, $a^k$ denotes $k$ repeated coordinates equal to $a$. For $a\leq t$, write $\clamp(z;a,t)=\min\{t,\max\{a,z\}\}$.

For an allocation $x$ at peaks $p$, a coordinate with $x_i<p_i$ is a shortage and one with $x_i>p_i$ is a surplus. LEFT and RIGHT refer to these two cases for the reporting agent, and fulfillment means $x_i=p_i$. In the proof, OWN abbreviates the unilateral strategy-proofness condition~\eqref{eq:sp}.

\subsubsection{Feasible transfers and global block capacities}
\label{u26:I-1}

\begin{lemma}
\label{u26:transfers-capacities}
A transfer \(\delta\) from coordinate \(i\) to coordinate \(j\), which leaves
its donor no lower than its recipient, is feasible. For \(1\leq j\leq n-1\),
let \(M_j\) be the maximum sum over \(j\) designated coordinates. It is
positive and independent of the designated labels, and
\[
 \left(\left(\frac{M_j}{j}\right)^j,
 \left(-\frac{M_j}{n-j}\right)^{n-j}\right)
\]
is feasible. Every \(j\)-slot sum is at most \(M_j\), and every complementary
\((n-j)\)-slot sum is at least \(-M_j\).

At every relative-boundary point \(b\), write \(a=\min b\), \(t=\max b\);
let \(J\) be the strictly intervening positions, and let \(\lambda,\nu\) be
the common averaged support coefficients on the minimum and maximum blocks.
Then \(C=\nu-\lambda>0\), \(A_i=\nu-\rho_i\in[0,C]\), and
\begin{equation}
 \sum_{i\in J}A_i(x_i-b_i)+C\sum_{b_j=a}(x_j-a)\geq0
 \quad\text{for every }x\in X .
 \tag{SUP}\label{u26:SUP}
\end{equation}
\end{lemma}

\begin{proof}
The transfer is a convex mixture of the vector and its \((i,j)\)-coordinate
swap, so it is feasible. A finite matching of donor excesses to recipient
deficits is feasible whenever each final donor target is at least each final
recipient target: before each step the donor is no lower than its target, the
recipient no higher than its target, and one remaining excess or deficit is
exhausted.

Compactness attains \(M_j\), and symmetry makes it independent of labels. It
is positive because the \(j\) largest coordinates of a nonzero zero-sum
vector have positive sum. Average a maximizer within the block and within
its complement to get the displayed feasible endpoint ray. The capacity
bounds follow. They hold at every feasible allocation.

For the support certificate, average a proper \(H_0\)-supporting
normal \(\rho\) at \(b\) over all equal-coordinate blocks. Since \(0\) is
relatively interior, its value at \(b\) is positive; averaging preserves that
value and properness. If \(b_i<b_j\), the feasible swap gives
\[
 (\rho_i-\rho_j)(b_j-b_i)\leq0,
\]
so the averaged coefficients are ordered. The gap is positive, for otherwise
all coefficients are equal and hence zero in \(H_0\). Eliminating
the maximum block using budget gives \eqref{u26:SUP}. Every water tie,
including filled central caps, belongs to the full extreme groups.
\end{proof}

\subsubsection{Uniform gains as capped maximal sign volume}
\label{u26:I-2}

\begin{lemma}
\label{u26:capped-volume}
On the peak box at \(p\), \(U\) is the unique leximin optimum and is the
capped vector at the maximum feasible sign volume. It is total, peak-only,
feasible, and physically symmetric.
\end{lemma}

\begin{proof}
For sign class \(I_s\), with caps \(a_i=|p_i|\), the unique capped gain vector
at volume \(V\) is
\[
 g_i(V)=\min\{a_i,c_s(V)\},\qquad \sum_i g_i(V)=V.
\]
The scalar \(c_s\) is the common uncapped gain level on a sign side, referred to below as its water level. It need not be unique when all caps are filled, but the gain
vector is. Zero peaks have coordinate zero. Capped vectors are continuous
and coordinatewise nondecreasing in \(V\). On the fixed peak box, sign
volume is the linear functional
\[
 V(x)=\sum_{p_i>0}x_i.
\]

Any feasible peak-box vector at volume \(V\) can be made capped at that same
volume. On a sign side, an excessive gain relative to \(g(V)\) has final
target \(c_s\), while a deficient gain has final target at most \(c_s\).
Match excesses to deficits by the noncrossing transfers of
Lemma~\ref{u26:transfers-capacities}; negative gains reverse coordinate signs
but preserve the same noncrossing averages. Thus the capped \(x(V)\) is
feasible, and its feasibility witnesses that volume. The intersection of
\(X\) and the peak box is compact, convex, and contains \(0\). Its
linear-volume image is \([0,V^*]\), with an attained maximum; the preceding
canonicalization makes \(x(V)\) feasible at every such volume.

Each nontrivial same-sign canonicalizing average strictly improves the sorted
absolute-gain vector: it raises a smaller gain without crossing the larger one
or violating a cap. Thus \(x(V)\) lexicographically dominates every other
peak-box vector at that volume. Increasing feasible \(V\) weakly increases
every capped gain and strictly some. Consequently \(x(V^*)\) is exactly the
unique defining leximin optimum \(U\). If either sign class is empty, balance
forces zero. In particular, a feasible capped vector filling all caps of one
sign is \(U\).

The construction exists at every peak vector, uses only peaks, and chooses a
point of \(X\). A simultaneous permutation of physical coordinates and
peaks preserves the peak box and sorted absolute-gain objective. Uniqueness
therefore permutes \(U\), proving permutation symmetry on the complete report domain. Translation and whole-environment
reflection conjugate the optimizer. Individual strategy-proofness and Pareto efficiency are proved below.
\end{proof}

\begin{lemma}
\label{u26:capped-boundary}
For legal peaks \(p\), suppose
\(b_i=\clamp(p_i;a,t)\) is feasible, with \(a<0<t\), and with nonempty full
extreme groups
\[
 E_-=\{i:b_i=a\},\qquad E_+=\{i:b_i=t\}.
\]
Put \(J=(E_-\cup E_+)^c\). Then
\begin{align}
 U_i&=p_i=b_i &&(i\in J),\nonumber\\
 U_i&\leq a &&(i\in E_-),\nonumber\\
 U_i&\geq t &&(i\in E_+).
 \tag{ENV}\label{u26:ENV}
\end{align}
If \(b\) is boundary, then \(U=b\).
\end{lemma}

\begin{proof}
At \(V_b=\sum_i\max\{b_i,0\}\), \(b=x_p(V_b)\), so \(V^*\geq V_b\). Every
\(J\)-cap is already filled, so capped-gain monotonicity gives
\eqref{u26:ENV}. Zero peaks remain zero. Full extreme groups include
water-tied central caps, some of which may be saturated and remain fixed. If
\(b\) is boundary, use \eqref{u26:SUP}, with \(\lambda,\nu\) on the full
extreme groups and \(C=\nu-\lambda>0\). For
\(\Delta=V^*-V_b\), fixed \(J\)-coordinates account for all other sign
volume:
\[
 \sum_{E_+}(U_i-b_i)=\Delta,\qquad
 \sum_{E_-}(U_i-b_i)=-\Delta,
\]
and hence
\begin{equation}
 0\geq\rho\cdot(U-b)=C\Delta\geq0.
 \tag{CB}\label{u26:CB}
\end{equation}
Thus \(\Delta=0\), and capped-vector uniqueness gives \(U=b\).
\end{proof}

For canonical \(U\), either at least one sign has every cap filled, or both
signs have unfilled caps. In the latter case both signs are nonempty, and
relative interior of \(0\) supplies a small feasible positive/negative trade
in the peak box, so \(V^*>0\). All unfilled positive coordinates are at the
positive maximum of \(U\), and all unfilled negatives at its negative minimum.
Then \(U\) must be boundary: otherwise increasing one such positive coordinate
and decreasing one such negative coordinate by the same small amount raises
feasible peak-box volume. The first alternative includes empty sign classes.
In particular, strict unfilled literal \(L/H\) waters place the canonical allocation on the boundary.

\subsubsection{Two canonical cap-persistence rules}
\label{u26:I-3}

\begin{lemma}
\label{u26:nesting-relaxation}
If a new peak box is a subset of the old one and still contains its canonical
\(u\), then the canonical vector is the same whole \(u\):
\begin{equation}
 U_{\rm new}=u.
 \tag{NEST}\label{u26:NEST}
\end{equation}
Tied caps and simultaneous contractions are allowed. If one same-sign cap
\(a_i=|p_i|>|u_i|\) is relaxed to \(a_i'\geq a_i\), the canonical allocation
is unchanged. This may be repeated for finitely many strict relaxations.
\end{lemma}

\begin{proof}
For nesting, use the fixed leximin objective and uniqueness. Under a strict
relaxation the old box lies in the new one, so \(V'\geq V\). If \(V'>V\), mix
\(u\) with a new maximizing vector at a sufficiently small positive weight.
Every unchanged coordinate satisfies its old cap, and the changed coordinate
has strict old slack. The mixture lies in the old peak box, is feasible by
convexity, and has sign volume exceeding \(V\), a contradiction. Hence
\(V'=V\). At that volume the relaxed capped vector remains \(u\), because
its old water is below both caps.
\end{proof}

\subsubsection{Canonical strategy-proofness}
\label{u26:I-4}

\begin{lemma}
\label{u26:canonical-median}
Fix other peaks and put
\(\lambda=U_i(\ell,p_{-i})\), \(\eta=U_i(\bar p,p_{-i})\). For every legal
peak \(p\),
\begin{equation}
 U_i(p,p_{-i})=\med\{\lambda,p,\eta\}.
 \tag{U-MED}\label{u26:U-MED}
\end{equation}
The canonical rule is individually own-strategy-proof on the full original
domain.
\end{lemma}

\begin{proof}
The endpoint peak boxes give \(\lambda\leq0\leq\eta\). At \(p=0\), the
coordinate is zero. For \(p>0\), write \(u_i=U_i(p,p_{-i})\in[0,p]\). If
\(\eta\leq p\), the \(H\)-endpoint canonical vector belongs to the peak-\(p\)
box, a subset of the \(H\)-box; \eqref{u26:NEST} gives \(u_i=\eta\). If
\(\eta>p\) and \(u_i<p\), relaxation of the strictly nonbinding cap to
\(\bar p\) gives \(\eta=u_i<p\), a contradiction. Thus \(u_i=p\). This
gives \(u_i=\min\{p,\eta\}\), including cap ties, zero volume, and
\(p=\bar p\). Reflection of the entire environment gives
\(u_i=\max\{p,\lambda\}\) for \(p<0\), proving
\eqref{u26:U-MED}.

At fixed opponent peaks, the range of the reporting agent's allocation is exactly \([\lambda,\eta]\), since
each peak in that interval is fixed. Every complete strictly single-peaked
true order uniquely prefers its median point to all other points of this
interval, including endpoint truths. Every alternative report returns a
point in the same interval. This proves strategy-proofness on the full
original domain.
\end{proof}

\subsubsection{Whole-\texorpdfstring{\(X\)}{X} efficiency of the canonical rule}
\label{u26:I-5}

\begin{lemma}
\label{u26:canonical-efficiency}
The canonical rule is Pareto efficient against all of \(X\).
\end{lemma}

\begin{proof}
Suppose \(y\in X\) weakly Pareto improves \(u=U(p)\), strictly for someone.
Filled peaks are fixed, including zero peaks. At an unfilled positive peak,
\(p_i>u_i\geq0\) forces \(y_i\geq u_i\); at an unfilled negative peak,
\(p_i<u_i\leq0\) forces \(y_i\leq u_i\). These directions permit crossing
peaks and cross-wing indifference; \(y\) need not remain in the peak box.
Budget gives
\[
 \Delta=\sum_{p_i>0}(y_i-u_i)
       =\sum_{p_i<0}(u_i-y_i)>0.
\]
If either sign is wholly filled, its sum is zero, a contradiction. Otherwise
the capped-volume characterization puts \(u\) on the boundary with all
unfilled coordinates in full extreme groups. Apply \eqref{u26:SUP}. Every
other coordinate, and every filled cap tied at an extreme, is fixed.
Consequently
\[
 \rho\cdot(y-u)=\nu\Delta-\lambda\Delta=C\Delta>0,
\]
contradicting global support. This proves Pareto efficiency on \(X\).
\end{proof}

\subsection{Attainable shares, efficient transfers, and endpoint profiles}
\label{u26:II}

The actual-menu classification is proved in Lemma~\ref{u26:actual-menus} in the main text.

\subsubsection{CUT and a fixed-peak flip}
\label{u26:II-2}

\begin{lemma}
\label{u26:CUT}
At an actual efficient allocation, there cannot be agents $i,j$ with $x_i>p_i$, $x_j<p_j$, and $x_i>x_j$.
If a peak-\(p\) report receives \(x<p\) and has an actual own offer \(b>p\) at
the same opponents, some full peak-\(p\) report receives a surplus.
\end{lemma}

\begin{proof}
If a strict surplus exceeds a strict shortage, a sufficiently small
donor-to-recipient transfer avoids both peaks and crossing the two
coordinates, strictly improving both. We refer to this implication as CUT.

Choose a full peak-\(p\) report strictly preferring \(b\) to \(x\). If its
new quantity \(y\) were at most \(p\), original-truth strategy-proofness
would force \(y\leq x\), while under the new truth \(b\) beats \(y\). Thus
the new report is surplus.
\end{proof}

\subsubsection{Literal endpoints and arbitrary marked replacements}
\label{u26:II-3}

\begin{lemma}
\label{u26:endpoint-source}
At \(L^l,H^h\), with \(l,h\geq1\) and \(l+h=n\), the actual and canonical row
is
\[
 E=(\alpha^l,\beta^h),\qquad
 \alpha=-\frac{M_h}{l}<0<\beta=\frac{M_h}{h}.
\]
Write these values \(\alpha_{(l,h)},\beta_{(l,h)}\). Replacing marked lows
by independent complete reports with peaks at most \(\alpha\), and marked highs by
independent complete reports with peaks at least \(\beta\), leaves the whole actual
row equal to \(E\).
\end{lemma}

\begin{proof}
Actual symmetry gives two quantities. A nonnegative low amount is improved
by a small positive endpoint trade. A negative low and positive high
determine a ray; actual efficiency forces its maximal endpoint \(E\). The subscripts record the two block sizes. Canonically \(E\) is capped and attains maximal volume \(M_h\), so
\(U=E\) directly. All-\(L\) and all-\(H\) rows are zero by symmetry and
budget, also canonically by the peak box.

Induct on marked replacements, allowing literal no-ops. For a marked low,
its own return to \(L\) is \(\alpha\) by the smaller-mark hypothesis. True
\(L\) makes \(\alpha\) a minimum; a true peak at most \(\alpha\) strictly
prefers \(\alpha\) to every larger amount, pinning that agent. A marked high
is symmetric: its own \(H\)-target gives maximum \(\beta\) and pins it. Every
other complete report stays fixed in each comparison.

If residual literal blocks have amounts \(c,k\), budget gives
\[
 l_0(c-\alpha)+h_0(k-\beta)=0.
\]
A missing block is handled by budget. Otherwise the original \(h\) high
slots contain \(h-h_0\) pinned \(\beta\)'s and \(h_0\) literal \(k\)'s.
Their global capacity gives
\[
 (h-h_0)\beta+h_0k\leq M_h=h\beta,
\]
so \(k\leq\beta\) and \(c\geq\alpha\). Feasible \(E\) fixes every marked
agent, lowers \(L\), and raises \(H\); current-row efficiency forces equality.
This covers independent shapes, both sides, threshold ties, and endpoint
no-ops. Canonical nesting independently gives \(E\) after cap contractions.
We denote the endpoint allocation result by EB and the replacement result by HS.
\end{proof}

\subsubsection{Weighted shortage and exact MIN}
\label{u26:II-4}

\begin{lemma}
\label{u26:weighted-shortage}
At an actual row suppose
\[
 \sum_{i\in J}A_i(x_i-b_i)\geq0,\qquad A_i\geq0.
\]
Fix \(j\) with \(A_j>0\). For every other positive-weight \(i\), require
either an independent actual upper bound \(x_i\leq b_i\), or true peak
\(p_i\leq b_i\) and \(b_i\geq b_j\). Then
\[
 x_j\geq\min\{p_j,b_j\}.
\]
Zero coefficients impose no condition. A sliced inequality requires omitted
coordinates to have been independently pinned at that actual row.
\end{lemma}

\begin{proof}
Under a contrary shortage, CUT bounds every surplus in the second group by
\(x_j<b_j\leq b_i\), and nonsurpluses are already at most \(p_i\leq b_i\).
All terms are nonpositive and the \(j\)-term is strictly negative, a
contradiction.
\end{proof}

\begin{lemma}
\label{u26:MIN-lemma}
On a fixed-peak product of full-report classes, suppose
\begin{equation}
 \sum_{i\in J}A_i(F_i-p_i)\geq0,\qquad A_i>0,
 \tag{MIN}\label{u26:MIN}
\end{equation}
throughout. Each indexed agent has an actual offer below \(p_i\) at
every tuple, and its class contains a left selector whenever that menu omits
\(p_i\). Then every indexed agent receives her peak throughout the product.
\end{lemma}

\begin{proof}
Otherwise pick a witness \(j\), at the least peak, with a violation somewhere
in the product. Every smaller indexed peak is pinned at every tuple. Omit
those zero terms: weighted shortage, with \(b_i=p_i\geq p_j\) on remaining
indices, gives \(F_j\geq p_j\) throughout. Thus the witness is surplus and
its peak is absent by strategy-proofness. Its permitted left selector at
the same opponents stays in the product and contradicts that bound. An empty
indexed set is vacuous.
\end{proof}

\subsection{A full-cylinder completion principle}
\label{u26:III}\label{u26:iii}

\begin{lemma}
\label{u26:full-cylinder}
Distinguish \(m\) central positions, \(k\) fixed complete low reports
\(T_1,\ldots,T_k\) with peaks at most \(d\), and \(h\) literal \(H\) reports,
where \(m+k+h=n\) and \(h\geq1\). Let central reports range over their full
fixed-peak cylinder. Fix a feasible benchmark
\[
 u=(u_1,\ldots,u_m,d^k,t^h)
\]
and a set \(J\) such that \(u_i=p_i\) for \(i\in J\), while
\(F_i=u_i\) throughout this entire cylinder for every other central. Every
\(J\)-agent has an actual \(H\)-offer above \(p_i\) at every complete
opponent tuple. If every physical low amount \(y_\ell\geq d\) throughout the
cylinder, then \(F=u\).
\end{lemma}

\begin{proof}
If \(\sum_{i\in J}(F_i-p_i)>0\), budget and low bounds force common literal
\(H\)-amount \(c<t\). Feasible \(u\) Pareto improves the row: it fixes pinned
centrals, fills all \(J\)-peaks, weakly lowers lows, and strictly raises \(H\).
Therefore
\[
 \sum_{i\in J}(p_i-F_i)\geq0
\]
throughout. Apply weighted MIN in the whole reflected environment, with unit
positive weights on \(J\). The \(H\)-offers become strict left offers,
and full classes provide every selector. Thus all \(J\)-agents are fulfilled.

Budget gives
\[
 \sum_\ell(y_\ell-d)+h(c-t)=0.
\]
Feasible \(u\) fixes centrals and improves every different completion, so
efficiency gives every \(y_\ell=d\) and \(c=t\). With \(k=0\), use budget;
\(J=\varnothing\) is included. Reflect the whole environment for the dual.
With no endpoint block, a feasible benchmark fulfilling every unpinned agent
and fixing the others Pareto-completes the row directly.
\end{proof}

\subsubsection{Completion with endpoint bounds and fixed coordinates}
\label{u26:III-1}

\begin{lemma}
\label{u26:local-oe-osa}
If each interior's actual \(L/H\)-target is canonically identified uniformly
in a fixed-peak full cylinder, then actual menus pin nonopen agents; open
agents have \(u_i=p_i\) and actual \(H\)-offers above \(p_i\). The \(H\)-only
cylinder is then completed by Lemma~\ref{u26:full-cylinder}; reflect for
\(L\)-only, and use its final direct observation for no endpoints.

More generally, on an \(H\)-only cylinder suppose a fixed set \(E\) is already
pinned to \(u\), every other interior has \(u_i\leq p_i\), and its actual
\(H\)-return is canonically identified. Then the cylinder is identified.
\end{lemma}

\begin{proof}
The first statement follows by applying Lemma~\ref{u26:full-cylinder}
with the specified endpoint allocations. For the
general statement, if \(\eta_i=u_i<p_i\), the canonical median gives its
actual \(H\)-maximum, and same-wing strategy-proofness pins it. If
\(\eta_i=u_i=p_i\), it is an attained peak. Remaining agents have
\(u_i=p_i<\eta_i\), so full-cylinder completion applies once. The set \(E\)
may be empty. The fixed coordinates and \(H\)-endpoint allocations hold at every tuple of complete reports.
\end{proof}

\subsection{Nonnegative support and the exact endpoint-bound bridge}
\label{u26:IV}

\subsubsection{Actual clipping from weighted shortage}
\label{u26:IV-1}

\begin{samepage}
\begin{lemma}
\label{u26:actual-clipping}
If \(p_j\leq b_j\) and weighted-shortage hypotheses hold for every complete report
at that peak with those exact opponents, then
\[
 F_j=\max\{p_j,K\},
\]
where \(K\) is its actual \(L\)-minimum. At an ordered boundary benchmark
\(b\), if every position outside the maximum block has true peak no greater
than its benchmark coordinate, \eqref{u26:SUP} excludes shortages at any
minimum-block position and gives its fixed-menu clipping formula.
\end{lemma}
\end{samepage}

\begin{proof}
For \(p_j<K\), the minimum is best on the right wing. For \(K\leq p_j\), an
omitted peak admits a full left selector contradicting the no-shortage bound.
At \(p_j=\ell\), equality is the minimum identity itself. In the boundary
application, maximum-block reports can be arbitrary: their coordinates were
eliminated using the common support coefficient of the maximum block.
\end{proof}

\subsubsection{Identification from two endpoint allocations}
\label{u26:IV-2}\label{u26:iv:projection}

\begin{lemma}
\label{u26:direct-ebd}
Fix a whole fixed-peak cylinder
\((R_1,\ldots,R_m,L^l,H^h)\), where \(l,h\geq1\). Suppose every interior's
actual \(L/H\)-return is canonical at every full opponent tuple. Write
\(U=u=(u_i,d^l,t^h)\), and suppose actual \(F_L=c\leq d\) throughout the
cylinder. Then \(F=u\) throughout; the whole-environment dual holds with
\(F_H\geq t\).
\end{lemma}

\begin{proof}
Actual menus pin all nonopen interiors. Let \(J\) be remaining agents, with
\(u_i=p_i\) and strict actual left/right offers. The capped construction gives
\(d\leq p_i\leq t\) for \(i\in J\). A small positive endpoint trade and
capped volume give \(d<0<t\). If \(d=\ell\), feasibility gives \(c\geq d\),
and full-cylinder completion applies using the strict \(H\)-offers. If \(t=\bar p\), use the dual with actual left offers.
Thus the projection-endpoint cases follow directly from feasibility and the endpoint allocations.

Retain \(\ell<d<0<t<\bar p\). The canonical \(u\) is boundary; otherwise
decreasing \(L\) and increasing \(H\) by a small equal total, fixing finite
coordinates, raises peak-box volume. Apply full-stabilizer support, setting
\(A_i=\nu-\rho_i\) for every physical interior, including water ties. After
omitting already pinned interiors, every actual row obeys
\begin{equation}
 \sum_{i\in J}A_i(F_i-p_i)+lC(c-d)\geq0.
 \tag{SUP-EBD}\label{u26:IV-SUP}
\end{equation}
The bound \(c\leq d\) and \eqref{u26:IV-SUP} give \eqref{u26:MIN} for positive-weight
\(J\)-agents on the entire full cylinder. Each has a left minimum
below \(p_i\) and full left selectors, so weighted MIN pins them all at once.
Equation~\eqref{u26:IV-SUP} and \(c\leq d\) then give \(c=d\). Zero-weight agents
retain \(u_i=p_i\) and their \(H\)-offers above \(p_i\); full-cylinder
completion closes the entire allocation, including either empty weight group. Reflect
the entire environment for the dual.
\end{proof}

\subsection{Induction with one low-endpoint report}
\label{u26:V}\label{u26:v}

\begin{samepage}
\begin{lemma}
\label{u26:singleton-induction}
Let \(S(m,l,h)\) denote whole canonical identification at that actual
signature, universally over legal interior peaks and independent full shapes,
in the same fixed \(n\)-agent environment. If \(m,h\geq1\), \(m+1+h=n\), and
\begin{equation}
 S(m-1,2,h),\qquad S(m-1,1,h+1),\qquad S(m,0,h+1),
 \tag{INPUT}\label{u26:INPUT}
\end{equation}
then \(S(m,1,h)\) follows.
\end{lemma}
\end{samepage}

\begin{proof}
The first two families supply every original interior's actual \(L/H\)-anchor;
the third supplies auxiliary \(H\)-targets. Write
\(U=(u_i,d,t^h)\). Use strong induction over
\[
 k=\#\{i:u_i>p_i\},
\]
uniformly over all level-\(m\) peak vectors and full shapes at this signature.
The projection-endpoint argument of Lemma~\ref{u26:direct-ebd} handles \(d=\ell\) or
\(t=\bar p\). Otherwise \(\ell<d<0<t<\bar p\). The exceptional set \(J\)
has \(p_j<d\), \(u_j=d\); all other negative interiors are fulfilled and all
nonnegative coordinates are at most their peaks.

If an original actual \(L\)-amount is \(z>d\), change only \(L\) to a full
report \(T_d\) with peak exactly \(d\), producing an \(H\)-only auxiliary
\(Q\) with \(m+1\) interiors. Its peak box is a subset of the original and
contains \(u\). Nesting gives
\begin{equation}
 U(Q)=u,\qquad U_T(Q)=d,\qquad J(Q)=J.
 \tag{WATER}\label{u26:WATER}
\end{equation}
Every
other auxiliary interior has \(u_i\leq p_i\) unless it belongs to \(J\).

For each \(j\in J\), the negative cap is strictly nonbinding:
\(p_j<d=u_j\). Relax \(j\) to literal \(L\); strict cap relaxation preserves
the canonical allocation. The actual target has signature \((m,1,h)\) and
exactly \(k-1\) exceptions, since new \(T_d\) is fulfilled at \(d\). Strong
induction identifies it for every retained complete report. True \(L\) makes
returned \(d\) the menu minimum; the true lower peak \(p_j\) strictly prefers
\(d\) to every larger amount. Own-strategy-proofness pins just \(j\) to \(d\)
on \(Q\)'s entire fixed-peak full-shape cylinder, including every independent
\(T_d\)-report. If \(J=\varnothing\), no smaller-\(k\) target is needed.

Every other auxiliary interior's actual \(H\)-target has signature
\((m,0,h+1)\), known from \eqref{u26:INPUT} at every retained complete tuple.
Lemma~\ref{u26:local-oe-osa} identifies \(Q\), using the fixed coordinates in \(J\). Its
\(T\)-quantity \(d<z\) is a profitable unilateral report by original true
\(L\), a contradiction. This proves the upper bound \(F_L\leq d\) on the entire original cylinder;
the two endpoint allocations from \eqref{u26:INPUT} allow
Lemma~\ref{u26:direct-ebd} to identify the allocation. Strong \(k\)-induction is complete. Reflect the whole environment
for singleton-high.

The auxiliary \(m+1\)-interior profile is identified from the
smaller-\(k\) induction hypothesis and the \(H\)-endpoint allocations
in \eqref{u26:INPUT}, using Lemma~\ref{u26:local-oe-osa}.
\end{proof}

The common induction step derived from this lemma is stated and proved in Section~\ref{u26:v:ladder}.

\section{Endpoint configurations for five and six agents}\label{app:endpoints}

\subsection{One-reporter frontiers and canonical \texorpdfstring{\(H\)}{H}-thresholds}
\label{u26:VI}\label{u26:vi}

\subsubsection{Geometry for every \texorpdfstring{\(k+h=n-1\)}{k+h=n-1} signature}
\label{u26:VI-1}

\begin{lemma}
\label{u26:one-reporter-frontier}
Fix \(k,h\geq1\), \(k+h=n-1\), in physical order
\((\text{reporter},k\text{ lows},h\text{ highs})\). Put
\[
 A=-\frac{M_h}{k+1},\qquad C_{\max}=\frac{M_h}{h},\qquad
 B=\frac{M_{h+1}}{h+1},\qquad D_{\min}=-\frac{M_{h+1}}{k}.
\]
For \(q\in[A,B]\), let \(C(q)\) maximize \(c\) in a feasible
\((q,d^k,c^h)\), where \(q+kd+hc=0\), and put
\[
 D(q)=\frac{-q-hC(q)}{k},\qquad V(q)=(q,D(q)^k,C(q)^h).
\]
Then
\begin{equation}
 \begin{gathered}
 D_{\min}\leq D(q)\leq A,\qquad B\leq C(q)\leq C_{\max},\\
 D(A)=A,\quad C(A)=C_{\max},\quad D(B)=D_{\min},\quad C(B)=B.
 \end{gathered}
 \tag{G}\label{u26:G}
\end{equation}
Moreover, \(D\) is nonincreasing with the strict-drop certificate below.
\end{lemma}

\begin{proof}
The feasible endpoints are
\[
 E_A=(A,A^k,C_{\max}^h),\qquad E_B=(B,D_{\min}^k,B^h).
\]
Thus \(A=\alpha_{(k+1,h)}<0<B=\beta_{(k,h+1)}\). Applying \(h\)-slot and
\(k\)-slot capacities to \(E_B,E_A\) gives
\(B\leq C_{\max}\), \(D_{\min}\leq A\). The endpoint chord makes every
fiber nonempty; compactness attains the maximum. The same chord and
capacities give \eqref{u26:G}: the chord's \(d\leq A,c\geq B\), and maximizing
\(c\) lowers \(d\). At \(A\) high capacity is attained; at \(B\), low
capacity and budget force \(c\leq B\). Thus
\(D(q)\leq q\leq C(q)\), strictly inside \((A,B)\), with opposite-sign waters.
\(V(q)\) is relative boundary: otherwise increase highs and offset lows.

For \(A\leq x<y<B\), mix \(V(x)\) with \(E_B\) and maximize the high fiber at
\(y\). Then
\[
 D(y)\leq\frac{(B-y)D(x)+(y-x)D_{\min}}{B-x}\leq D(x),
\]
strictly when \(D(y)>D_{\min}\), since then \(D(x)>D_{\min}\). At \(y=B\),
use \(D(B)=D_{\min}\).

At a feasible row with reporter \(s\in[A,B]\), average designated blocks.
Their means satisfy \(\bar d\geq D(s)\), \(\bar c\leq C(s)\). The condition
\(s\in[A,B]\) holds, in particular, for an ordered feasible \((s,d^k,c^h)\), \(d\leq s\leq c\): for \(s<0\), reporter-plus-low mean is between \(A\) and \(s\); for
\(s>0\), reporter-plus-high mean is between \(s\) and \(B\), by global
capacities. At zero it is automatic.
\end{proof}

\subsubsection{Canonical endpoint bounds and the scalar identity}
\label{u26:VI-2}

\begin{lemma}
\label{u26:pure-envelope}
Let
\[
\begin{gathered}
 b=(p,r_1,\ldots,r_m,a_1,\ldots,a_k,t^h)\in X,\\
 1+m+k+h=n,\quad h\geq1,\quad t>0,\quad p\leq t,\quad a_j\leq0.
\end{gathered}
\]
Retained central peaks are \(r_i\), retained low peaks satisfy
\(\theta_j\leq a_j\), and designated highs have literal peak \(\bar p\). At
the canonical \(H\)-endpoint of the first agent,
\(\eta=U_i(H,p_{-i})\) satisfies \(\eta\geq p\), strictly if \(p<t\).

For every legal reporter peak \(p\), put \(q=\med\{A,p,B\}\). Then
\begin{equation}
 U(R_p,L^k,H^h)=V\bigl(\med\{A,p,B\}\bigr).
 \tag{U}\label{u26:U}
\end{equation}
\end{lemma}

\begin{proof}
For \(p<0\), \(\eta\geq0\). For \(p\geq0\), \(b\) belongs to enlarged peak
box, whose maximal volume is at least
\[
 V_b=p+ht+\sum_i r_i^+.
\]
The \(h+1\) literal \(H\)-coordinates have water \(\eta\), including complete
positive saturation, so
\[
\begin{gathered}
 V=g(\eta),\qquad g(s)=(h+1)s+\sum_i\min\{r_i^+,s\},\\
 V_b-g(p)=h(t-p)+\sum_i[r_i^+-\min\{r_i^+,p\}]\geq0.
\end{gathered}
\]
The right side is positive if \(p<t\). Strict increase of \(g\) on
\([0,\bar p]\) proves the claim, including \(p=0\). Reflection by
\(x\mapsto-x\) gives the corresponding strict lower-endpoint bound above a
negative low water.

At the endpoint profiles, the canonical reporter receives \(A,B\); \eqref{u26:U-MED} gives the reporting agent amount \(q\). Canonical
literal blocks are equal and maximize high fiber: a feasible larger high at
fixed \(q\) lowers its low, stays in the same peak box, and has greater sign
volume, a contradiction. The frontier identifies \eqref{u26:U}. This formula includes thresholds, exterior peaks, and projection endpoints.
\end{proof}

\subsubsection{A bound on the allocation after a high-endpoint report}
\label{u26:VI-3}

\begin{lemma}
\label{u26:H-threshold}
At fixed physical positions, let
\[
 b=(p,r_1,\ldots,r_m,a_1,\ldots,a_k,\underbrace{t,\ldots,t}_{h\text{ times}})
 \in X,
\]
where \(1+m+k+h=n\), \(h\geq1\), \(t>0\), \(p\leq t\), and \(a_j\leq0\).
The reporter's true peak is \(p\); retained central reports have peaks \(r_i\);
retained low reports have peaks \(\theta_j\leq a_j\); and the \(h\) designated
high reports are literal \(H\). If the reporter's actual change to \(H\) at
these \(n-1\) complete opponents is canonically identified, its allocation \(\eta\) satisfies
\begin{equation}
 \eta\geq p;\qquad \eta>p\quad\text{if }p<t.
 \tag{H-THRESHOLD}\label{u26:H-THRESHOLD}
\end{equation}
\end{lemma}

\begin{proof}
By hypothesis, the allocation after the report \(H\) agrees with its
canonical value at the fixed opponent reports. Lemma~\ref{u26:pure-envelope}
shows that this value is at least \(p\), and strictly greater when \(p<t\).
\end{proof}

\subsection{A finite raising family for two-low signatures}
\label{u26:VII}\label{u26:vii}

\begin{center}
\(\displaystyle
 (n;m,k,h)\in\{(5;1,2,2),(6;1,2,3),(6;2,2,2)\},\qquad
 D=(p_1,\ldots,p_m,a^k,t^h),\quad k=2 .
\)
\end{center}

Equivalently, the one-central case has \(m=1\), \(k=2\),
\(h\in\{2,3\}\), and \(n=h+3\); the two-central case has \(n=6\) and
\(m=k=h=2\). Fix a feasible boundary benchmark
\[
 \ell<a<\min\{p_1,\ldots,p_m,0\},\qquad
 \max\{p_1,\ldots,p_m,0\}<t<\bar p.
\]
The common geometric input is a finite raising family: for some \(K\geq1\),
\(\Lambda=K+1\), and \(\eta_0>0\), every central \(i\) and
\(0<\eta\leq\eta_0\) have the feasible vector
\begin{equation}
 D_i(a-\eta)=\left(p_1,\ldots,p_i+\Lambda\eta,\ldots,p_m,
 a-\eta,a-\eta,\left[t-\frac{(\Lambda-2)\eta}{h}\right]^h\right).
 \tag{RAISE}\label{u26:RAISE}
\end{equation}
All constants and strict order/projection radii precede reports. Under a strict endpoint drop, the same
pure canonical \(H\)-endpoint chord supplies the family for every listed
signature. In the one-central applications, Section~\ref{u26:ix:frontier}
later identifies the endpoint with its global block-capacity vector, and
Section~\ref{u26:ix:BT} treats the no-drop case.

Average a proper support over lows and highs. Then \(C=\nu-\lambda>0\),
\(0\leq A_i=\nu-\rho_i\leq C\), and
\begin{equation}
 \sum_iA_i(x_i-p_i)+C\sum_{\rm low}(x_j-a)\geq0.
 \tag{S}\label{u26:S}
\end{equation}
Substituting \eqref{u26:RAISE} gives \(\Lambda A_i\geq2C\), so every
\(A_i>0\). Actual clipping yields LOW at all independent low tuples with
peaks at most \(a\) and every full central tuple: low agent \(j\)'s amount is
\(\max\{\text{peak},\kappa_j\}\), where \(\kappa_j\) is that agent's exact
actual \(L\)-minimum at the fixed opponents.
Maximum reports may be arbitrary for LOW; all actual sources below use
literal \(H\).

\subsubsection{The canonical \texorpdfstring{\(H\)}{H}-endpoint and its chord}
\label{u26:VII-1}\label{u26:vii:benchmark}

\begin{lemma}
\label{u26:pure-endpoint-chord}
Relabel the central peaks so that \(q=p_m\) is greatest. At the pure
canonical profile retaining \(p_1,\ldots,p_{m-1}\), with two \(L\) reports
and \(h+1\) \(H\) reports, write
\[
 E=(p_1,\ldots,p_{m-1},c,b,b,c^h),\qquad
 b\leq a,\quad c\geq \frac{q+ht}{h+1}>\max\{q,0\}.
\]
If \(b<a\), the endpoint chord supplies the finite raising family
\eqref{u26:RAISE}. If \(b=a\) in the two-central six-agent case, it instead
supplies a proper support at \(D\) whose high and \(q\)-coefficients agree,
whose high--low gap is positive, and whose retained central coefficient lies
between the low and high coefficients. Section~\ref{u26:viii:two} uses this
plateau support. In the one-central applications, the no-drop case is
completed separately by the global-capacity argument in
Section~\ref{u26:ix:BT}.

From any supplied raising vector, transferring \(\Lambda\eta\) from the raised
central to either low yields the feasible partner in either low orientation,
with the same high water.
\end{lemma}

\begin{proof}
Relabel centrals so \(q=p_m\) is greatest. Average it with the \(h\) highs
of \(D\). The feasible trial has retained peaks \(p_i\), two lows \(a\), and
\(h+1\) equal highs
\[
 \sigma=\frac{q+ht}{h+1}>\max\{q,0\},
\]
by order and budget. At the pure canonical profile with other central peaks
retained, two \(L\)'s, and \(h+1\) \(H\)'s, the capped envelope gives
\[
 E=(p_1,\ldots,p_{m-1},c,b,b,c^h),\qquad b\leq a,\quad c\geq\sigma>q.
\]
Use raw support at \(D\), allowing \(A_m=0\) before rays. If \(b<a\),
substitution of feasible \(E\) gives
\[
 A_m(c-q)\geq2C(a-b).
\]
Since \(A_m\leq C\), put
\(\Lambda=(c-q)/(a-b)\geq2\), \(K=\Lambda-1\). For
\(\tau=\eta/(a-b)\in(0,1)\), the chord
\((1-\tau)D+\tau E\) is exactly \(D_m(a-\eta)\). Budget gives
\[
 c-q=2(a-b)+h(t-c),
\]
so \(c\leq t\) and high water is \(t-(\Lambda-2)\eta/h\). Shrink
\(\eta_0\) for every strict order/projection margin before reports. For
\(p_i<q\), also require \(p_i+\Lambda\eta<q\) and transfer
\(\Lambda\eta\) from raised \(q\) to \(i\). The donor ends at \(q\), above
recipient, yielding \(D_i(a-\eta)\) with lows and highs unchanged. For
\(p_i=q\), swap central coordinates instead. Thus the same family works for
equal and zero peaks as well.

For the two-central six-agent signature, write \(p=p_1\). If \(b=a\), budget
gives \(c=\sigma\), so \(E\) is exactly \(D\) averaged over \(\{q,H,H\}\).
Both literal waters of \(E\) are strictly inside projection caps, so canonical
\(E\) is boundary. Average a proper support there over equal-coordinate
blocks, with normal \(\rho\). Its value \(\rho\cdot E\) is positive; the
three \(c\)-coefficients are equal, hence \(\rho\cdot D=\rho\cdot E>0\).
The same normal properly supports \(D\), with positive high/low gap,
\(q\)/high coefficients equal, and retained \(p\)-coefficient between low and
high. This is the plateau support used in Section~\ref{u26:viii:two}.
Each one-central no-drop case is handled separately by the global-capacity
argument in Section~\ref{u26:ix:BT}.

From a raising vector, transfer \(\Lambda\eta\) from raised central to either
low. The donor ends at \(p_j\); recipient ends at
\[
 (a-\eta)+\Lambda\eta=a+K\eta<p_j,
\]
by shrunk radius. The noncrossing transfer gives the partner, in either low
orientation, with high water \(t-(K-1)\eta/h\).
The transfer applies also at \(K=1\). All radii precede reports.
\end{proof}

\subsubsection{Peak attainment and an affine comparison family}
\label{u26:VII-2}\label{u26:vii:water}

\begin{samepage}
\begin{lemma}
\label{u26:filled-partner}
Assume \(S(m+1,0,h+1)\) holds for all
legal peaks, independent complete reports, and physical placements. If
\[
 b=(r_1,\ldots,r_{m+2},s^h)\in X,\qquad
 \ell<r_j<s,\quad0<s<\bar p,
\]
then at every independent full tuple at true interior peaks \(r_j\) and
literal \(H^h\), \(b\) is the entire allocation:
\begin{equation}
 F(R_1,\ldots,R_{m+2},H^h)=b,\qquad p(R_j)=r_j.
 \tag{FILLED}\label{u26:FILLED}
\end{equation}
For \(v=a-\eta\), \(0<\eta\leq\eta_0\), the transferred partner is
\begin{equation}
 T_v=(p_1,\ldots,p_m,v,u(v),t(v)^h),\qquad
 u(v)=a+K(a-v),\quad t(v)=t-\frac{(K-1)(a-v)}{h}.
 \tag{P}\label{u26:P}
\end{equation}
For every independent full central tuple and both independent full low
reports at the displayed peaks \(v,u(v)\), with literal \(H^h\), the whole
actual row is \(T_v\), in either low orientation.

For \(a-\eta_0\leq z<a\), fix the full central reports, literal \(H^h\), and
one complete low report \(V_z\) with peak \(z\). If \(K_z\) is the other
low agent's literal-\(L\) minimum, that agent's return \(c_z\) at any full
report with peak \(z\) satisfies
\[
 z\leq c_z=\max\{z,K_z\}\leq a+K(a-z).
\]
At low peaks \(a,a\), the entire allocation is \(D\) for every full central
tuple and both independent full peak-\(a\) reports, with literal \(H^h\).
This is the WATER row; its corresponding literal-\(L\) minimum satisfies
\(K_a\leq a\).
\end{lemma}
\end{samepage}

\begin{proof}
Each physical interior's own \(H\)-target has required prior signature at
exact retained reports. The threshold certificate gives an attainable share
above \(r_j\). Full-cylinder completion, with all interiors active and no
fixed low positions, identifies the entire allocation.

For \eqref{u26:P},
\[
\begin{gathered}
 2a-v\leq u(v),\qquad \ell<v<a<u(v)<\min\{p_1,\ldots,p_m,0\},\\
 \max\{p_1,\ldots,p_m,0\}<t(v)\leq t<\bar p.
\end{gathered}
\]
All displayed interiors are new true peaks below positive high water, so
FILLED identifies whole \(T_v\) for every full central tuple and both
independent low reports. Both \(u(v)\to a\) and \(t(v)\to t\) follow from
the affine formulas. The required family is \(S_n(2,0,h+1)\) for \(m=1\),
where \(n=h+3\) and \(h\in\{2,3\}\), and \(S_6(3,0,3)\) for \(m=2\).
These are \(H_2\) and \(H_3\), respectively.

Fix full centrals, a complete \(V_z\) of peak \(z<a\), and literal \(H^h\).
The other low's change from \(L\) to complete report of peak \(u(z)\) returns
\(u(z)\). Its literal-\(L\) minimum \(K_z\leq u(z)\). Together with LOW,
\[
 z\leq c_z=\max\{z,K_z\}\leq a+K(a-z).
\]
At low peaks \(a,a\), feasible \(D\) also satisfies
FILLED, giving the whole WATER row for all full centrals and independent full
peak-\(a\) reports. Thus \(K_a\leq a\).
\end{proof}

\subsubsection{Attainable shares from the raising family}
\label{u26:VII-3}\label{u26:vii:raise}

\begin{lemma}
\label{u26:actual-raise}
Assume the identified family and the FILLED conclusion of
Lemma~\ref{u26:filled-partner}. Let \(z=a-\eta\) with
\(0<\eta\leq\eta_0\), within the same strict order and projection
neighborhood. Hold all other complete central reports fixed at the chosen
original reports with peaks \(p_j\). Hold the two chosen complete low reports
with peak \(z\), together with literal \(H^h\), fixed throughout the deviation.
At these same complete opponents, a complete report with peak
\(p_i+\Lambda\eta\) returns exactly \(p_i+\Lambda\eta\). Consequently,
\begin{equation}
 F_i\leq p_i+\Lambda\eta
 \tag{UP}\label{u26:UP}
\end{equation}
for every full original central tuple.
\end{lemma}

\begin{proof}
The vector has same high water \(t(z)\) as partner, since \(\Lambda=K+1\).
The chosen radius keeps every displayed interior peak in projection and
below its positive adjusted high water, itself below \(\bar p\). FILLED
therefore identifies the whole actual raising row at new true peaks, using the
same prior \(H\)-family at every retained tuple. The conclusion includes equal peaks and peaks crossing zero.

At original copy opponents, central \(i\) can report peak
\(p_i+\Lambda\eta\) and receive it. Own-strategy-proofness gives
\eqref{u26:UP}. At mixed low reports, these vectors serve as feasible Pareto comparisons. All strict sign, cap, and
projection margins precede report choices. The two-central six-agent branch uses the canonical boundary benchmarks
proved separately in Section~\ref{u26:xii}. The one-central branches use
the frontier benchmarks of Section~\ref{u26:ix:frontier} and the prior
source of Section~\ref{u26:ix:H2}.
\end{proof}

\subsubsection{Cofinal completion using the same finite rays}
\label{u26:VII-4}\label{u26:vii:tail}

\begin{lemma}
\label{u26:cofinal-completion}
Let \(z\uparrow a\). For each \(z<a\) sufficiently close to \(a\), fix
complete low reports \(V_z,W_z\), each with true peak at most \(z\).
Fix this entire family independently of the central report choices.
Let centrals range independently
over fixed full-report classes \(\mathcal C_i\) at \(p_i\). Each has an actual \(H\)-offer in \((p_i,\bar p]\) at every tuple, and \(\mathcal C_i\)
contains, for every legal \(\varepsilon>0\), a report preferring \(\bar p\) to
\(p_i-\varepsilon\), fixed independently of \(z\). Suppose that for every
fixed class tuple, eventually
\begin{equation}
 \text{both actual low amounts }y_1,y_2\text{ are at least }z.
 \tag{TAIL}\label{u26:TAIL}
\end{equation}
The tail may depend on that tuple, including a fixed selector replacement.
Then the entire actual row tends to \(D\) for every fixed tuple.
\end{lemma}

\begin{proof}
For each fixed tuple, label every central by its
peak and both lows by \(a\). If convergence fails, budget supplies a central
or low witness. Choose the greatest label admitting a nonvanishing deviation at any
fixed class tuple. A selected low has excess at least \(\varepsilon\) above
\(a\) on a subsequence, by \eqref{u26:TAIL}. A selected central shortage
\(x_i\leq p_i-\varepsilon\) is converted by one fixed class report preferring
\(\bar p\) to \(p_i-\varepsilon\). Every \(H\)-offer in
\((p_i,\bar p]\) is then preferred to the shortage, so the fixed-peak flip
makes this same report right throughout the subsequence. Continuity of the
old order at its peak and strategy-proofness keep the right return a fixed
distance away. Replacement remains in class, and TAIL reapplies at its tail.

At the fixed tuple/subsequence put \(\eta=a-z\). For a central witness use
feasible \(D_i(z)\); for low witness orient feasible partner to that low.
Each vector fills other centrals, puts unselected lows at \(z\), and adds
\(\Lambda\eta\) to selected central's \(p_i\) or selected low's \(z\). The
latter target is \(z+\Lambda\eta=u(z)\). Since selected quantity has fixed
excess, and true low peak is at most \(z\), at most actual low, every
non-\(H\) agent weakly improves eventually. Both vectors have same high
water \(t(z)\), with gain
\begin{equation}
 h(t(z)-k_{\rm actual})
 =\sum_{\rm central}(x_i-p_i)+\sum_{\rm low}(y_\ell-z)-\Lambda\eta.
 \tag{GAIN}\label{u26:GAIN}
\end{equation}

For a central witness, CUT excludes shortages at lower/tied central peaks;
higher labels converge at this tuple by maximality. Selected surplus
contributes a fixed positive amount. For low witness, all central labels are
higher and converge, while its own low summand stays positive; every other low
summand is nonnegative. Thus GAIN is eventually positive, contradicting
efficiency. All central and low coordinates converge; literal symmetry and
budget finish.
\end{proof}

\subsection{Identifying allocations with repeated low reports}\label{u26:VIII}\label{u26:viii}

Use the finite family of \S\ref{u26:vii} at one of its three listed
\(k=2\) signatures, together with the identified \(H\)-only family
for that signature. Require \(\mathsf{COPY\text{-}LEFT}\) at \emph{every}
full central tuple and every repeated full low report \(V,V\) with
\(p(V)<a\): the actual \(L\) return of each changing central is strictly
below its peak. Assume that at every independent strict-low tuple \(V,W\) with both peaks
below \(a\), each central agent's actual \(H\)-return is strictly above her peak. These are separate returns at exact
\((n-1)\)-opponent tuples, including every subsequently fixed selector
replacement.

The conclusion is the whole \(D\) for every independent full central tuple
and every strict copy \(V,V\) with \(p(V)<a\); in particular, it includes
literal \(L,L\). Water copies with peak \(a\) use
\S\ref{u26:vii:water} separately. Each one-central application uses
\(\mathsf{HS}\) and \(S_n(2,0,h+1)\). The two-central, six-agent application uses the allocation at three identical
low reports, denoted \(\mathsf{TP}\) and established in Lemma~\ref{u26:x:TP},
together with \(S_6(3,0,3)\).

\subsubsection{Rational preference classes and repeated low reports}
\label{u26:viii:rational-contract}

For each central peak \(b\), fix the class
\[
 \mathcal C_b=\{R_b^k:k\in\mathbb Q_{>0}\},
\]
where \(R_b^k\) is the complete report represented on the whole projection by the following loss function, with lower losses preferred:
\[
 \ell_b^k(x)=
 \begin{cases}
   (b-x)^2,&x\leq b,\\
   k(x-b),&x\geq b.
 \end{cases}                                                   \tag{Q}\label{u26:viii:Q}
\]
These are continuous strictly single-peaked complete reports. They provide the following two preference comparisons. If a compact actual menu omits
\(b\) and has adjacent offers \(l<b<u\), any rational
\[
 k>\frac{(b-l)^2}{u-b}
\]
makes its LEFT offer strictly best. For a fixed \(\varepsilon>0\) with
\(b-\varepsilon\in I\), any rational
\[
 0<k<\frac{\varepsilon^2}{\bar p-b}
\]
makes the report strictly prefer the physical RIGHT endpoint to that
shortage. The bounds are strict and finite, so the rational choice can be
made separately at each physical central peak. The second report is fixed
independently of any later actual \(H\) offer.

This family is fixed independently of the menus and low reports.
All comparisons in \eqref{u26:viii:Q} hold on the full projection. Actual
menu compactness and its attained neighboring pair are from
\S\ref{u26:ii:menus}.

For a proper subset \(E\) of the centrals, let \(B\) be its nonempty
complement and put
\[
 \mathcal C_E
 =
 \prod_{i\in E}\mathcal A_{p_i}
 \times
 \prod_{i\in B}\mathcal C_{p_i},
\]
where \(\mathcal A_b\) is the full class of reports with peak \(b\).
Assume \(\mathsf{COPY\text{-}PIN}\): every \(E\)-agent receives precisely
its peak throughout \(\mathcal C_E\) at every strict copy \(V,V\). The argument below identifies the entire allocation \(D\) at these
profiles and gives every \(B\)-agent her peak for all complete reports
with that peak, at the same opponents. In the convergence argument,
all central reports vary: the \(E\)-agents range over their full classes.

\subsubsection{Bounds for rational preferences at repeated low reports}
\label{u26:viii:quantitative}

Fix a cofinal family of full \(V_z\), \(z<a\), independently of all later
central-report choices, and write \(\eta=a-z\). Copies have actual common
low \(c_z\) and
\begin{equation}
 c_z=\max\{z,K_z\},\qquad z\leq c_z\leq a+K\eta . \label{u26:viii:copy-bound}
\end{equation}
The actual \(\mathsf{RAISE}\) offers of \S\ref{u26:vii:raise} give
\(\mathsf{UP}\) for every original full central tuple. For each test agent
\(i\in B\), set
\[
 M_i=\frac{2CK+\Lambda\sum_{j\in B,\;j\ne i}A_j}{A_i}.
\]
At every tuple of \(\mathcal C_E\), \((S)\), exact
\(\mathsf{COPY\text{-}PIN}\), \eqref{u26:viii:copy-bound}, and the
\(\mathsf{UP}\) bounds for the other test agents give
\begin{equation}
 F_i\geq p_i-M_i\eta. \label{u26:viii:lower}
\end{equation}
Fix a tuple of \(\mathcal C_E\). If test agent \(i\) is RIGHT, its actual
menu omits \(p_i\) by \(\mathsf{OWN}\). The COPY-LEFT condition and
\S\ref{u26:viii:rational-contract} supply the nearest actual LEFT offer
\(l_i\), selected inside its rational class at the same opponents. The
bound \eqref{u26:viii:lower} applies afresh at the replacement, whence
\[
 p_i-l_i\leq M_i\eta.
\]
The original true parameter \(k_i\) and \(\mathsf{OWN}\) give
\[
 k_i(F_i-p_i)\leq(p_i-l_i)^2\leq M_i^2\eta^2.
\]
Thus
\[
 F_i\leq p_i+\frac{M_i^2}{k_i}\eta^2;
\]
the bound is automatic on LEFT. With
\[
 E_0=\sum_{i\in B}\frac{A_iM_i^2}{2Ck_i}>0,
\]
support now yields
\begin{equation}
 c_z\geq a-E_0\eta^2>z,
 \qquad\text{and hence}\qquad K_z=c_z\longrightarrow a, \label{u26:viii:K-limit}
\end{equation}
on the geometric tail \(\eta<1/E_0\). The strict inequality identifies the menu minimum as \(c_z\). This applies to every fixed tuple of \(\mathcal C_E\),
including every later fixed RIGHT replacement in a full \(E\)-class or a
rational \(B\)-class, with its own \(E_0\) and tail. All \(E\)-pins used
here concern copies and hold on the entire product.

\subsubsection{Convergence with two different low reports}
\label{u26:viii:reference-ceiling}

Fix any full strict reference \(V_0\) with peak \(v_0<a\). At a fixed
tuple of \(\mathcal C_E\), let \(K_0\) be the actual \(L\) minimum opposite
\(V_0\). It is fixed as \(z\) varies. On the tail \(z>v_0\),
\eqref{u26:viii:K-limit} and the two physical crossed \(\mathsf{LOW}\)
identities give
\begin{equation}
 \text{mixed lows}
 =
 \bigl(K_z,\max\{z,K_0\}\bigr),
 \qquad K_z>z. \label{u26:viii:mixed-lows}
\end{equation}
At every fixed tuple of \(\mathcal C_E\), on its own tail, both true low
peaks are at most \(z\) and both actual low amounts are at least \(z\).
Apply \(\mathsf{TAIL}\) from \S\ref{u26:vii:tail} with all centrals active,
using the full classes for \(E\) and the rational classes for \(B\). Both
contain its fixed RIGHT selectors, and \(H\) supplies every agent's actual
opposite offer. Every fixed replacement remains in \(\mathcal C_E\), where
the bounds for repeated low reports apply again.
\(\mathsf{TAIL}\) gives whole mixed convergence to \(D\). Hence
\(K_0\leq a\), since a fixed \(K_0>a\) would prevent the second low from
tending to \(a\). Because the strict reference was arbitrary, both lows are
at most \(a\) at every \(\mathcal C_E\) copy.

\subsubsection{Identification at repeated low reports}
\label{u26:viii:complete-copies}

On the COPY product, \(\mathsf{COPY\text{-}PIN}\) omits the \(E\)-agents,
and the low ceilings from \S\ref{u26:viii:reference-ceiling} turn \((S)\)
into \((\mathsf{MIN})\) for the positive \(B\)-weights. COPY-LEFT gives
every \(B\)-agent a strict LEFT offer at every retained tuple, and
\S\ref{u26:viii:rational-contract} supplies its LEFT selector inside the
rational class. Weighted MIN pins all \(B\)-peaks. Then \((S)\), the two
ceilings, and budget give the whole copy \(D\). Each attained \(B\)-peak
separately pins its global full role at those exact COPY opponents by
\(\mathsf{OWN}\). This proves the conditional identification result.

\subsubsection{Extending from rational to arbitrary central preferences}
\label{u26:viii:induction}

Prove the conditional identification result for every proper \(E\) by induction on
\(\lvert E\rvert\). For \(E=\varnothing\),
\(\mathsf{COPY\text{-}PIN}\) is vacuous. If \(E\ne\varnothing\) and
\(i\in E\), the preceding assertion for \(E\setminus\{i\}\) has \(i\)
rational. Its conclusion for arbitrary reports at repeated low opponents pins \(i\) under any full own
report, while all other \(E\)-reports are arbitrary complete reports and the
remaining \(B\)-reports stay rational. Thus \(i\) is pinned throughout
the entire \(\mathcal C_E\) on every copy. Doing this separately for each
\(i\) gives precisely \(\mathsf{COPY\text{-}PIN}\), so the conditional identification result applies.

For an arbitrary full central tuple, apply the conclusion for arbitrary reports at repeated low opponents
with \(E\) equal to all agents except \(i\), separately for every \(i\).
All central amounts are pinned on every strict copy. There, \((S)\), low
\(\mathsf{SYM}\), and budget give \(c\geq a\) and high at most \(t\).
Feasible \(D\) Pareto-completes the entire allocation because the copied true peak is
below \(a\). Hence all full central tuples have whole \(D\) on all strict
copies.

\subsubsection{The low-endpoint profile}\label{u26:viii:literal}

Since \(\ell<a\), literal \(L,L\) is a strict copy. The preceding induction
therefore identifies the whole \(D\) here for every independent full central
tuple. This conclusion is used in Section~\ref{u26:ix} for one interior report
and in the six-agent argument below for two interior reports. Each
application at a nearby or reflected benchmark requires the endpoint
allocations and raising family specified at the start of this subsection.

\subsubsection{Boundary identification with two interior reports}
\label{u26:viii:two}

Let \(n=6\). Consider a feasible boundary
\[
 D=(p,q,a,a,t,t),\qquad a<p<q<t,\qquad a<0<t,
\]
with strict projection waters. At every full \(Q_q\), the actual
\(P\to L\) return is below \(p\). At every full \(P_p\), the actual
\(Q\to H\) return is above \(q\). Assume that the preceding argument identifies
every two-central benchmark with strict endpoint drop \(b<a\) in
\S\ref{u26:vii:benchmark}, using
\S\ref{u26:viii:rational-contract}--\S\ref{u26:viii:literal} in both whole
orientations. This hypothesis includes nearby canonical benchmarks. Then the allocation at the endpoint profile is \(D\) for all full central reports.
Denote this implication by \(\mathsf{TWO}\). Projection cases use the independent
two-endpoint identification argument of \S\ref{u26:iv:projection}.

If the endpoint in \S\ref{u26:vii:benchmark} strictly drops, apply the
identification result for repeated low reports. Otherwise its plateau support at \(D\) has equal
\(q\)- and high-coefficients. Normalize its positive high--low gap; the
remaining \(p\)-coefficient lies between them. Thus
\[
 x_{L_1}+x_{L_2}+\phi x_P\geq 2a+\phi p,
 \qquad 0\leq\phi\leq1. \tag{LF}\label{u26:viii:LF}
\]
For sufficiently small \(\varepsilon>0\), raise \(p\) to
\(p+\varepsilon<q\) by mixing \(D\) with its two swaps of \(P\) and a high,
each with weight \(\varepsilon/[2(t-p)]\). The feasible trial has low
water \(a\) and high water \(t-\varepsilon/2\), retaining strict
central--water order. The canonical row at these central peaks and
\(LLHH\) fills both centrals by \eqref{u26:ENV}; write its
waters \(c_\varepsilon,h_\varepsilon\). \(\mathsf{ENV}\), budget, and the
old global inequality \eqref{u26:viii:LF} give
\[
 a-\frac{\phi\varepsilon}{2}\leq c_\varepsilon\leq a,
 \qquad
 h_\varepsilon=t-\frac{\varepsilon}{2}-(c_\varepsilon-a),
 \qquad
 t-\frac{\varepsilon}{2}\leq h_\varepsilon\leq t.
\]
One small geometric \(\varepsilon\)-interval preserves all strict
order and projection margins. These canonical rows are boundary rows
because literal water caps remain unfilled. At any unit normal there,
write \(\lambda,\nu\) for low/high averages and \(C=\nu-\lambda>0\).
Comparing old feasible \(D\) in this new support gives
\begin{equation}
 (\rho_P-\lambda)\varepsilon\geq2C(t-h_\varepsilon). \label{u26:viii:new-support}
\end{equation}

If \(h_\varepsilon<t\), the whole reflected benchmark cannot have an
endpoint plateau: \S\ref{u26:vii:benchmark} would supply a proper normal
with \(\rho_P=\lambda\), contradicting
\eqref{u26:viii:new-support}. Its reflected endpoint therefore strictly
drops, and the reflected identification result for repeated low reports applies. Its
offer to \(P\) \(p+\varepsilon\) is attainable at every full \(Q_q\), with
literal \(LLHH\) and \(Q\) fixed.

If \(h_\varepsilon=t\), the bounds force \(\phi=1\). Budget turns
\eqref{u26:viii:LF} into
\[
 x_Q+x_{H_1}+x_{H_2}\leq q+2t.
\]
The segment from \(D\) to this canonical row supplies feasible vectors
\[
 \bigl(Z,q,[a-(Z-p)/2]^2,t,t\bigr),
 \qquad p\leq Z\leq p+\varepsilon.
\]
Choose one full \(P_p\) that strictly prefers \(\ell\) to \(p+\varepsilon\),
independently of the full shape of \(Q\). Every actual \(L\) offer lies in
\([\ell,p)\), so \(\mathsf{OWN}\) gives its actual \(X<p+\varepsilon\) at
every \(Q\). If \(Q\)'s quantity \(Y>q\), \(\mathsf{CUT}\) forces
\(X\geq p\). The global bound makes actual high \(k<t\) and actual low
\(c\geq a-(X-p)/2\). The feasible segment vector at \(Z=X\) fixes \(P\),
improves \(Q\) to its peak, weakly lowers \(L\), and raises \(H\), contrary
to \(\mathsf{EFF}\). Thus every full \(Q\) is weakly LEFT at this fixed
\(P\). Its actual \(H\) offer above \(q\) and full-shape \(\mathsf{OWN}\)
force peak attainment, hence \(Q=q\) throughout this role. Now \(k\leq t\);
if \(X<p\), feasible \(D\) improves \(P,L,H\) while fixing \(Q\). Therefore
\[
 p\leq X<p+\varepsilon.
\]

For any \(\varepsilon_j\downarrow0\), either branch supplies offers to \(P\) in \([p,p+\varepsilon_j]\) at each fixed full
\(Q_q,L,L,H,H\). Closedness of that same actual menu attains \(p\), and
\(\mathsf{OWN}\) pins \(P\) for every full \(P\) at every such \(Q\).
Finally, \eqref{u26:viii:LF} gives the literal-low floor \(c\geq a\) on the
entire central cylinder. Section~\ref{u26:iii}, with \(Q\)'s \(H\) offer and the separately pinned \(P\), identifies the whole \(D\).

\subsection{One interior report at five and six agents}
\label{u26:IX}\label{u26:ix}

Fix \(h\in\{2,3\}\) and \(n=h+3\). Every capacity, feasible vector,
complete profile, and unilateral comparison in this section belongs to this
same \(n\)-agent environment. With the families \(H_j,B_j,A_j\) of
\S\ref{u26:v:ladder}, the target is
\[
 A_1=S_n(1,2,h).
\]
Use the fixed-\(n\) results of Sections~\ref{u26:I}--\ref{u26:VIII}: capacities and transfers
from Section~\ref{u26:I-1}; canonical volume, nesting, endpoints, and the median
from Section~\ref{u26:I-2}--Section~\ref{u26:I-4}; \(\mathsf{EB}\) and full
marked-replacement \(\mathsf{HS}\) from Section~\ref{u26:II-3}; full-cylinder
completion from Section~\ref{u26:iii}; independent projection closure from
\S\ref{u26:iv:projection}; the singleton ladder of Section~\ref{u26:v};
the \(k=2\) frontier of Section~\ref{u26:vi}; and
\S\ref{u26:vii}--\S\ref{u26:viii:literal} with \(m=1,k=2\), the present
\(h\), and separately supplied prior source \(S_n(2,0,h+1)\).
\subsubsection{Establishing \texorpdfstring{\(H_2\)}{H2}}
\label{u26:ix:H2}

Endpoint identification and Lemma~\ref{u26:local-oe-osa} give \(H_1,B_0,A_0\). Apply
\S\ref{u26:v:ladder} at \(j=1\). It gives
\begin{equation}
 B_1=S_n(1,1,h+1),
 \qquad
 H_2=S_n(2,0,h+1),                                   \label{u26:ix:H2-source}
\end{equation}
universally over legal peaks, independent complete reports, and placements.

The endpoint specializations of \(H_2\) follow from the same families: one new \(L\)
gives \(B_1\), one new \(H\) gives \(H_1\), and two new endpoints give
\(\mathsf{EB}\). Reflected \(B_1\) and \(H_1\) are available by
reflection of the entire environment. In particular, the prior
source is \(S_5(2,0,3)\) at \(h=2\) and \(S_6(2,0,4)\) at \(h=3\).

\subsubsection{Frontier geometry and attainable endpoint shares}
\label{u26:ix:frontier}

Apply Section~\ref{u26:vi} with \(k=2\) and the present \(h\). Put
\[
 A=\alpha_{(3,h)}=-\frac{M_h}{3},
 \qquad
 C_{\max}=\frac{M_h}{h},
 \qquad
 B=\beta_{(2,h+1)}=\frac{M_{h+1}}{h+1},
 \qquad
 D_{\min}=-\frac{M_{h+1}}{2}.
\]
The pure endpoint vectors are
\[
 E_A=(A,A,A,C_{\max}^{\,h}),
 \qquad
 E_B=(B,D_{\min},D_{\min},B^{\,h}).
\]
For \(q\in[A,B]\), the maximal-high frontier of Section~\ref{u26:VI-1} is
\[
 V(q)=(q,D(q),D(q),C(q)^h),
\]
where
\[
\begin{gathered}
 D_{\min}\leq D(q)\leq A,\qquad B\leq C(q)\leq C_{\max},\\
 D(A)=A,\quad C(A)=C_{\max},\quad D(B)=D_{\min},\quad C(B)=B.
\end{gathered}
\]
The independently proved canonical identity Section~\ref{u26:VI-2} gives
\[
 U(R,L,L,H^h)=V\bigl(\operatorname{med}(A,p(R),B)\bigr) .\tag{IX-U}\label{u26:ix:U}
\]
Fix \(A<q<B\), write \(a=D(q)\) and \(t=C(q)\). Then
\[
 \ell\leq a<\min\{q,0\},
 \qquad
 \max\{q,0\}<t\leq\bar p,
 \qquad
 D_{\min}\leq a\leq A.
\]

At any independent complete low tuple \(V,W\) with peaks at most \(a\), the
reporter's change to \(L\) produces \((L,V,W,H^h)\).
\(\mathsf{HS}(3,h)\) applies because all three low peaks are at most
\(A=\alpha_{(3,h)}\): the literal \(L\) peak is \(\ell\leq A\), and the
other two are at most \(a\leq A\). Its returned amount is \(A<q\). Thus
this is an actual LEFT offer at every exact tuple, including every
repeated copy, selector replacement, and cofinal tuple used in
Section~\ref{u26:viii}.

The reporter's change to \(H\) at those same opponents has two independent
complete reports and \(h+1\) literal highs. The prior \(H_2\) source
\eqref{u26:ix:H2-source} identifies that whole target, including its
endpoint specializations. Section~\ref{u26:VI-3}, applied to feasible \(V(q)\),
gives an actual \(H\) offer above \(q\) at every such low tuple and
every reporter shape. The same \(H_2\) is exactly the required
\(S_n(m+1,0,h+1)\) at \(m=1\) for Section~\ref{u26:VII-2}--Section~\ref{u26:VII-3}.
It therefore gives \(\mathsf{FILLED}\) and the allocations in the affine,
\(\mathsf{WATER}\), and \(\mathsf{RAISE}\) families at their specified peaks,
including the \(H\)-deviation of either auxiliary low agent.

Suppose \(\ell<a\), \(t<\bar p\), and \(a>D_{\min}\). The pure canonical
\(H\) endpoint in Section~\ref{u26:VII-1} is \(E_B\), whose low coordinates
strictly drop. The finite raising family has
\[
 \Lambda=\frac{B-q}{a-D_{\min}}\geq2,
 \qquad
 K=\Lambda-1\geq1,
 \qquad
 \text{adjusted high water}=t-\frac{(\Lambda-2)\eta}{h}.
\]
Indeed, the two budget identities give
\[
 B-q=2(a-D_{\min})+h(t-B),
\]
and \(t\geq B\). Thus the bound holds at both \(h=2\) and \(h=3\).
The \(\mathsf{HS}\) return supplies COPY-LEFT; the prior \(H_2\) and
Section~\ref{u26:VI-3} supply \(H\) at every independent strict-low tuple.
The same \(H_2\) identifies the finite auxiliary profiles.
Sections~\ref{u26:viii:rational-contract}--\ref{u26:viii:literal}
therefore identify the whole
\[
 D=(q,a,a,t^h)
\]
for every full peak-\(q\) reporter and every strict repeated low copy.
Literal \(L,L\) is one such copy because \(\ell<a\).

The two-low support coefficients remain \(2C\); \(h\) is a positive block
count or divisor throughout this application. At \(h=2\), the two
capacities above are \(M_2\) and \(M_3\), both defined in the five-agent
environment.

\subsubsection{Boundary completion and direct endpoint restoration}
\label{u26:ix:BT}

At literal opponents \(L,L,H^h\), the reporter's actual \(L/H\) targets are
the \(\mathsf{EB}\) rows \(L^3H^h\) and \(L^2H^{h+1}\). The reporting agent receives \(A\) and \(B\). The independent two-endpoint identification argument of
\S\ref{u26:iv:projection} handles \(a=\ell\) or \(t=\bar p\).

Retain \(\ell<a<0<t<\bar p\). If \(a>D_{\min}\),
\S\ref{u26:ix:frontier} gives the entire allocation. If \(a=D_{\min}\), the global
complementary-block capacity gives
\[
 x_{L_1}+x_{L_2}\geq-M_{h+1}=2D_{\min}
\]
at every feasible row. The two literal \(L\) reports are identical, so
physical \(\mathsf{SYM}\) makes each actual low amount at least
\(D_{\min}=a\) throughout the complete reporter role. The actual
\(H\) offer \(B>q\) and Section~\ref{u26:iii} now give the whole \(V(q)\)
for every complete report at \(q\). This covers the no-drop case for both
values of \(h\).

For any complete \(R\) with \(p(R)\leq A\), \(\mathsf{HS}(3,h)\) restores it
as one marked low in \(E_A\) and gives the whole \(E_A=V(A)\). For
\(p(R)\geq B\), \(\mathsf{HS}(2,h+1)\) restores it as one marked high in
\(E_B\) and gives the whole \(E_B=V(B)\). Threshold equality, projection
endpoints, and literal no-ops are included in Section~\ref{u26:II-3}. Combining
these cases with the interior and projection cases yields
\[
 F(R,L,L,H^h)=V\bigl(\operatorname{med}(A,p(R),B)\bigr)
\]
for every full \(R\), every legal peak, and every physical placement. By
\eqref{u26:ix:U}, this is the canonical allocation. Therefore
\S\ref{u26:ix:H2}--\ref{u26:ix:BT} establish both
\[
 S_5(1,2,2)\qquad\text{and}\qquad S_6(1,2,3).
\]
Whole-environment reflection gives \(S_n(1,h,2)\). At \(h=2\) this has the
same count signature; at \(h=3\) it is \(S_6(1,3,2)\).

\subsubsection{The next step of the same ladder}\label{u26:ix:H3}

The new \(A_1\), earlier \(H_2\), and \(B_1\) now supply
\S\ref{u26:v:ladder} at \(j=2\). It yields
\begin{equation}
 B_2=S_n(2,1,h),
 \qquad
 H_3=S_n(3,0,h),                                      \label{u26:ix:H3-source}
\end{equation}
uniformly over legal peaks, independent complete reports, and placements.

For \(H_3\) endpoint specializations, one new endpoint gives \(B_2\) or
\(H_2\); two give \(A_1,B_1\), or \(H_1\); and three give \(\mathsf{EB}\).
These whole sources and their whole-environment reflections are available at
the retained complete reports.

At \(n=5\), this gives \(B_2=S_5(2,1,2)\) and \(H_3=S_5(3,0,2)\).
Reflection of \(B_2\) is \(A_2=S_5(2,2,1)\), so the \(j=3\) premises are
available and yield \(B_3=S_5(3,1,1)\) and \(H_4=S_5(4,0,1)\).

At \(n=6\), it gives \(B_2=S_6(2,1,3)\) and
\(H_3=S_6(3,0,3)\). Reflection of \(B_2\) gives \(S_6(2,3,1)\). The balanced family
\(A_2=S_6(2,2,2)\) is established in
Sections~\ref{u26:x}--\ref{u26:xiii} before the six-agent \(j=3\) step.
Within those six-agent sections, \(H3\) denotes this actual
\(S_6(3,0,3)\) family with the endpoint specializations just listed.

\section{The balanced six-agent configuration}\label{app:balanced}\label{u26:part:balanced}

Throughout this appendix, fix \(n=6\).
\(\mathsf{BT}\) denotes the \(h=3\) instance of \S\ref{u26:ix:BT};
\(\mathsf{H3}\) denotes \(S_6(3,0,3)\) with the endpoint specializations of
\S\ref{u26:ix:H3}. The five-agent proof is already complete in Section~\ref{u26:xiv:five}.

\subsection{Three identical low reports}\label{u26:X}\label{u26:x}

\subsubsection{The diagonal allocation and preliminary results}
\label{u26:x:permissions}

Use Section~\ref{u26:vi} with \(k=3,h=2\):
\[
\begin{gathered}
 V(s)=(s,D(s),D(s),D(s),C(s),C(s)),\qquad A\leq s\leq B,\\
 A=\alpha_{(4,2)}<0<B=\beta_{(3,3)},\qquad D_{\min}=-B.
\end{gathered}
\]
Write \(\mathsf{CAP3}\) for the global three-coordinate capacity bound
\(\sum_{\rm low}x_i\geq-M_3=3D_{\min}\) from
Lemma~\ref{u26:transfers-capacities}. Write \(\mathsf{AVG}\) for the
block-average bounds \(\bar d\geq D(s)\) and \(\bar c\leq C(s)\) at
each feasible allocation whose reporter receives \(s\in[A,B]\), as proved
in Lemma~\ref{u26:one-reporter-frontier}. That lemma also gives the
endpoint values and strict decrease of \(D\) above \(D_{\min}\). The actual
literal \(\mathsf{BT}\) source is the whole-environment dual of
\S\ref{u26:ix:BT}. \(\mathsf{H3}\) is \S\ref{u26:ix:H3} with its
classified endpoint specializations; \(\mathsf{HS}\) is
Section~\ref{u26:II-3}.

For \(A<q<B\), put \(a=D(q)\) and \(t=C(q)\).
\begin{lemma}\label{u26:x:TP}
For every complete reporter \(S_q\),
\[
 \mathsf{TP}:\qquad
 F(S_q,T,T,T,H,H)=V(q)
 \quad\text{for every full \(T\) with peak at most \(a\)}. \tag{TP}
\]
\end{lemma}
\begin{proof}
First consider the low agents' menus.

\subsubsection{Low-agent menu minima and a sufficient lower bound}
\label{u26:x:floor}

By Section~\ref{u26:II-1}, global actual menus are compact with \(L/H\) extrema.
For finitely many menus whose minima are below \(v\), use the full rational
report \(R_v^k\) of \eqref{u26:viii:Q}, with \(k\) exceeding every threshold
\((v-l_i)^2/(u_i-v)\) at adjacent offers \(l_i<v<u_i\). This same report
selects LEFT whenever \(v\) is omitted and selects \(v\) whenever it is
attained; menus with no right offer impose no bound. Choose \(k\) separately at each \(v\).

\(\mathsf{CUT}\) and Section~\ref{u26:IV-1}, applied to \(V(q)\), exclude low
shortages for all independent low peaks at most \(a\) and every complete reporter
of peak at most \(q\). For a fixed full \(S\) at \(q\), define
\[
 K_S(R,T)
 =\text{the actual \(L\) minimum at exact opponents }S,R,T,H,H.
\]
Physical \(\mathsf{SYM}\) gives \(K_S(R,T)=K_S(T,R)\) after relabeling.
At every triple of eligible full low reports, their separate actual menus
give
\[
 x_i=\max\{p_i,K_S(R_j,R_k)\},
 \qquad \{i,j,k\}=\{1,2,3\}. \tag{CLIP}\label{u26:x:CLIP}
\]
This identity applies separately to each low agent at the heterogeneous
profiles used below.

The reporter-\(L\) target is \(\mathsf{HS}(4,2)\), returning \(A\). Its
reporter-\(H\) target is the prior \(\mathsf{H3}\) source at the three
retained complete low reports. Applying Section~\ref{u26:VI-3} to feasible
\(V(q)\), with \(C(q)\geq B>q\), gives an actual \(H\) offer
\(\eta>q\). Both endpoint allocations hold at every retained low tuple
and for every complete reporter preference.

\noindent\textit{\(\mathsf{FLOOR} \).}
If all actual low amounts are at least \(a\) for every complete reporter at one
fixed low tuple, Section~\ref{u26:iii} gives the whole \(V(q)\): activate the
reporter, retain these exact complete low reports with peaks at most \(a\),
and use its \(H\) offer above \(q\).

\subsubsection{Identification of the diagonal allocation}
\label{u26:x:diagonal}

Fix \(A<q<B\), \(d=D(q)>D_{\min}\), and an arbitrary full \(T\) with peak
\(\theta\leq d\). We first identify the allocation at \(TTT\) as \(V(q)\). Let \(M_j\) be the four global reporter
menus at fixed opponents \(T^j,L^{3-j},H,H\). All have actual minimum
\(A\). Suppose \(q\) is absent from \(M_3\), and let \(a_0<q\) be its
largest actual offer below \(q\). Then \(J=(a_0,q)\) is a nonempty actual
omitted interval.

For each \(v\in J\), choose one full \(R_v\) making simultaneous LEFT choices
in \(M_1\) and \(M_3\), and use it unchanged in all four rows. Write
\(d_v=D(v)\). Then
\[
 \theta\leq d<d_v<D(a_0),
 \qquad
 A\leq r_1(v)\leq v.
\]
\(M_3\) returns \(a_0\). Its equal \(T\)-amounts are at least
\(D(a_0)>\theta\) by \(\mathsf{AVG}\); if they were larger, \(V(a_0)\)
would fix the reporter, lower all \(T\)-amounts toward their peak, and raise
\(H\). Thus the allocation at the terminal profile is \(V(a_0)\). \(\mathsf{BT}\) and the three actual low-agent
\(\mathsf{CLIP}\) edges give
\[
\begin{array}{ll}
\text{row 1}:&
\bigl(r_1(v),d_v,w(v),w(v),h_1(v),h_1(v)\bigr),\\
\text{row 2}:&
\bigl(r_2(v),e(v),e(v),D(a_0),h_2(v),h_2(v)\bigr),\\
&e(v)=\max\{\theta,w(v)\}.
\end{array}
\]
The retained \(L\) in row 2 is \(D(a_0)\), because its own change to \(T\)
returns \(D(a_0)>\theta\) in row 3.
\(\mathsf{AVG}\) at row 1 gives
\[
 d_v+2w(v)\geq3D(r_1(v))\geq3d_v,
\]
so \(w(v)\geq d_v>\theta\) and \(e(v)=w(v)\).

If \(r_2(v)\geq v\), \(V(v)\) fills the reporter's true peak, weakly lowers
both \(T\)-amounts, strictly lowers the literal-\(L\) amount
\(D(a_0)>d_v\), and raises each \(H\) by
\[
 \frac{r_2(v)-v+2w(v)+D(a_0)-3d_v}{2}>0.
\]
Hence \(r_2(v)<v\), so \(v\notin M_2\). This holds throughout \(J\), and
\(M_2\) misses \(J\); its LEFT return is a constant \(k_2\) there. In
row 2, both the reporter quantity \(k_2\) and literal-\(L\) quantity
\(D(a_0)\) are fixed. If \(w\) differed between two values of \(v\), the
actual vector with smaller \(w\) would improve both fixed \(T\) reports and
both \(H\) reports while fixing those other coordinates. The reporters'
true peaks differ, but their compared quantity is identical. Thus
\(\mathsf{EFF}\) makes \(w(v)=W\) constant on \(J\).

For each \(v\in J\), choose \(u\in(a_0,v)\). Constancy gives
\(W=w(u)\geq D(u)>D(v)\). If \(v\in M_1\), feasible \(V(v)\) would fix the
reporter and first \(T\), lower both literal-\(L\) amounts, and raise \(H\),
contrary to \(\mathsf{EFF}\). Thus \(M_1\) misses \(J\) entirely. Its LEFT
choice is a constant \(k_1\) throughout \(J\), and row 1 is
\[
 \left(k_1,D(v),W,W,
       \frac{-k_1-D(v)-2W}{2},
       \frac{-k_1-D(v)-2W}{2}\right).
\]
For \(v_1<v_2\) in \(J\), the actual vector at \(v_2\) lowers the first
\(T\)-amount strictly toward \(\theta\) and raises both \(H\)-amounts while
fixing reporter and literal-\(L\) quantities. It Pareto improves row 1 at
\(v_1\), a contradiction. Therefore \(q\in M_3\). Every full peak-\(q\)
reporter selects \(q\); \(\mathsf{AVG}\) and \(\mathsf{EFF}\) force all
\(T\)-amounts to be \(d\), and budget fixes \(H\). This proves the result on the strict \(D\)-branch for every full \(T/S\).

If \(A<q<B\) and \(d=D_{\min}\), then at \(TTT\), the three identical full low reports have one quantity \(c\).
Global \(\mathsf{CAP3}\) gives \(c\geq d\) at every complete reporter shape.
\(\mathsf{FLOOR}\) proves the whole \(V(q)\) for every full \(T\) with peak
at most \(d\), including peak \(d\). This establishes the diagonal allocation on the plateau.
\end{proof}

\subsubsection{Replacing a low-endpoint report at the water level}
\label{u26:x:promotion}

At \(A<q<B\), put \(d=D(q)\) and \(t=C(q)\). We establish \(\mathsf{PLL}\), the identity
\(F(S_q,P,L,L,H,H)=V(q)\) for every complete reporter \(S_q\) and
every complete \(P\) with peak \(p\leq d\), using \(\mathsf{TP}\),
\(\mathsf{BT}\), \(\mathsf{HS}\), \(\mathsf{H3}\), and
Section~\ref{u26:iii}. \(\mathsf{BT}\) gives \(P=d\) throughout the
original \((S_q,P,L,L,H,H)\) reporter role. If \(d=\ell\), then \(P=L\)
and \(\mathsf{BT}\) suffices. If \(d=D_{\min}\), \(\mathsf{CAP3}\) makes
the two identical literal lows at least \(d\) throughout this role, so
Section~\ref{u26:iii} finishes using the \(H\) offer from
\S\ref{u26:x:floor}. Retain \(\ell<d\) and \(d>D_{\min}\).

\noindent\textit{Water insertion.}
For any full \(T\) with peak \(\theta<d\) and any full \(W\) of peak \(d\),
the whole \((S_q,T,T,W,H,H)\) row is \(V(q)\). Let \(N,M\) be the fixed
reporter menus at \(TTW\) and \(TWW\), respectively.
\S\ref{u26:x:floor} supplies \(\mathsf{CLIP}\) throughout, \(N\)'s \(H\) offer above \(q\), and both reporter minima \(A\).

Suppose \(q\) is missing from \(N\), with adjacent offers \(a_0<q<b\).
For \(v\in(a_0,q)\), choose one full \(R_v\) selecting \(b\) in \(N\).
\(\mathsf{TP}\) at \(TTT\) and \(WWW\) separately exposes
\[
 K(T,T)=K(W,W)=D(v)>d.
\]
Thus \(N\)'s \(W\) and \(M\)'s \(T\) each receive \(D(v)\). The two
\(T\)-quantities in \(N\) are \(e=\max\{\theta,K\}\), where \(K\) is the
actual minimum opposite \(R_v,T,W,H,H\). If \(e\geq d\), \(V(q)\)
Pareto improves \(N\); hence \(K<d\). At the same \(R_v\), both own-allocation identities for the \(W\)-agents \(\max\{d,K\}=d\) make
\[
 M=(r,D(v),d,d,k,k).
\]
If \(r\geq q\), feasible \(V(q)\) fixes both \(W\)'s, improves or fixes the
reporter (whose true peak is \(v<q\)), lowers \(T\), and raises \(H\), since
\[
 2(t-k)=r-q+D(v)-d>0.
\]
Thus \(r<q\); \(\mathsf{HS}\) gives \(r\geq A\). If \(r\leq v\),
\(\mathsf{AVG}\) contradicts
\[
 \frac{D(v)+2d}{3}<D(v)\leq D(r).
\]
Hence \(v<r<q\). Start at any \(v_0\in(a_0,q)\) and call this actual
\(M\)-offer \(v_1\). At peak \(v_1\), another \(N\)-RIGHT selector would
similarly return above \(v_1\) in \(M\), although \(v_1\) is its attained
peak there, a contradiction. Hence \(q\) is attained in \(N\). For every
full \(S_q\), \(\mathsf{TP}/\mathsf{CLIP}\) at \(TTT\) exposes
\(K(T,T)=d\), hence \(W=d\); \(\mathsf{AVG}\) and \(\mathsf{EFF}\) give
the two \(T\)-amounts \(d\), and budget gives \(H=t\). This proves the claimed allocation at \(TTW\).

\noindent\textit{Identification of \(\mathsf{PLL}\).}
If \(p=d\), take \(T=L,W=P\) in water insertion. If \(p<d\), fix any
full \(W_d\). At \((S_q,P,L,W,H,H)\), \(P\)'s allocation after reporting \(L\) is
\(LLW\), already proved, so \(P=d\). The \(L\)-agent's change to the same
full \(P\) returns \(d\) at \(PPW\), also already proved.
\(\mathsf{CLIP}\) and \(p<d\) expose its actual minimum as \(d\). The
remaining \(W\) is at least \(d\), so Section~\ref{u26:iii} gives the whole
\(PLW\) row throughout the complete reporter role. Consequently a literal
\(L\) in the original \(PLL\) has offer \(d\) from reporting \(W\), and
its minimum is at most \(d\). Both literal lows are therefore at most \(d\),
while \(P=d\). A reporter return \(A\leq r<q\) would violate
\(\mathsf{AVG}\) and strict \(D\)-drop. The actual \(L\) offer \(A<q\) and
full LEFT selection force \(q\) to be attained; \(\mathsf{AVG}\), the two
low ceilings, and budget finish \(V(q)\).

Together, the report \(W\) with peak \(d\) handles \(p<d\) and \(p=d\),
and the block-capacity bound handles the plateau and projection endpoints.

\subsection{Two interior reports with repeated low reports}\label{u26:XI}\label{u26:xi}

\subsubsection{A lower offer from the diagonal allocation}\label{u26:xi:copy-left}

Fix
\[
 D=(p,q,a,a,t,t),\qquad p\leq q,
\]
satisfying the \(m=2\) benchmark conditions of Section~\ref{u26:vii} and
the strict endpoint drop of Section~\ref{u26:VII-1}. Its chord/transfer supplies
the finite family. At either changing central, write \(b\) for its true
peak and \(r\) for the retained central peak. Let \(\alpha>0\) be this
changing central's coefficient in the original global support, and put
\[
 A_3=\alpha_{(4,2)},\qquad B_3=\beta_{(3,3)}.
\]
All retained complete reports stay fixed in each actual return. The global
three-slot capacity and budget give \(r+2t\leq3B_3\); since \(t>r\),
\(r<B_3\).

If \(r\leq A_3\), four-slot capacity gives
\[
 b+r+2a\geq4A_3.
\]
Since \(a<r\leq A_3\), \(r+2a<3A_3\), hence \(a<A_3<b\). The changing
central's \(L\) target has four full low reports with peaks at most \(A_3\).
\(\mathsf{HS}(4,2)\) returns its own amount \(A_3<b\), including
\(r=A_3\).

If \(r>A_3\), average \((b,a,a)\) in \(D\). Section~\ref{u26:vi} gives
\[
 D_3(r)\leq\frac{b+2a}{3}<b,
 \qquad
 C_3(r)\geq t>r.
\]
Apply the original global support to that feasible frontier, in physical order
\[
 (D_3(r),r,D_3(r),D_3(r),C_3(r),C_3(r)).
\]
This gives
\[
 (\alpha+2C)(D_3(r)-a)\geq\alpha(b-a)>0.
\]
Thus \(a<D_3(r)<b\). For repeated complete low reports \(V,V\) of peak
\(v<a\), change the central to that same full \(V\). \(\mathsf{TP}\) at
\((R_r,V,V,V,H,H)\) gives the changing agent's offer \(D_3(r)\),
at exact opponents \(R_r,V,V,H,H\). Its literal-\(L\) minimum is therefore
at most \(D_3(r)<b\).

For \(H\), changing the central to \(H\) is the prior whole \(\mathsf{H3}\)
source from \S\ref{u26:ix:H3}, at the retained central and both exact low
reports. Section~\ref{u26:VI-3}, applied to \(D\), gives its real
\(\eta>b\). This holds at every independent strict-low tuple \(V,W\) and
every full central tuple, including all class selectors and cofinal tuples.
Endpoint substitutions use the prior \(\mathsf{H3}\) specialization list.
The same \(\mathsf{H3}=S(3,0,3)\) supplies the partner, \(\mathsf{WATER}\),
and \(\mathsf{RAISE}\) family of Section~\ref{u26:VII-2}--Section~\ref{u26:VII-3},
including each auxiliary low agent's \(H\) change.

Apply this argument separately to the two central agents, including \(p=q\).
The geometry of Section~\ref{u26:vii}, actual partner/\(\mathsf{RAISE}\)
offers, and the argument for repeated low reports in Section~\ref{u26:viii} identify the whole
\(D\) for all full central reports at every strict copy \(V,V\). Since
\(\ell<a\), literal \(L,L\) is complete, giving the actual \(\mathsf{STRICT}\)
input for Section~\ref{u26:xii}. \(\mathsf{WATER}\) handles peak-\(a\)
copies separately.

\subsection{All boundary cases, projection waters, and equal central peaks}
\label{u26:XII}\label{u26:xii}

The target is the whole literal \(LLHH\) row at any feasible boundary
\[
 D=(p,q,a,a,t,t),\qquad a<p\leq q<t,\qquad a<0<t.
\]
The \(\mathsf{STRICT}\) input is the literal conclusion of
\S\ref{u26:xi:copy-left}, established from the endpoint bounds and raising family for repeated low reports. \(\mathsf{BT}\)/dual is \S\ref{u26:ix:BT}, also in
every whole-environment reflection.

\eqref{u26:CB} identifies \(D\) canonically. Each central's
actual \(L/H\) targets are \(\mathsf{BT}\)/dual, universally in the other
complete report. Thus the independent projection closure of
\S\ref{u26:iv:projection} handles \(a=\ell\) or \(t=\bar p\), including
equal central peaks. Retain strict projection waters below.

\subsubsection{Endpoint allocations for arbitrary central preferences}
\label{u26:xii:anchors}

For a strict central pair \(p<q\) and strict projection waters, apply
Section~\ref{u26:VI-3} to \(D=(p,q,a,a,t,t)\), with \(Q\) as changing reporter
and \(P\) retained. Its actual \(Q\to H\) target is the prior
\(\mathsf{BT}\) profile \((P,L,L,H,H,H)\), canonically identified at every
full \(P\). Since \(q<t\), Section~\ref{u26:VI-3} gives \(Q\)'s \(H\)-report allocation above \(q\). Reflect the entire environment and apply the same
argument with \(P\) changing: prior dual \(\mathsf{BT}\) identifies its
actual \(L\) target \((L,Q,L,L,H,H)\), and \(a<p\) gives its \(L\)-report allocation below \(p\) at every full \(Q\).

\subsubsection{The two-central boundary completion}\label{u26:xii:two}

Section~\ref{u26:xii:anchors} supplies actual \(P\to L<p\) at every full
\(Q\) and actual \(Q\to H>q\) at every full \(P\). These are precisely the
endpoint bounds required by \(\mathsf{TWO}\) in \S\ref{u26:viii:two}.
Section~\ref{u26:xi:copy-left} supplies its strict input at every
strict-drop benchmark, including nearby canonical rows and throughout the
reflected environment.

The implication \(\mathsf{TWO}\) therefore identifies the whole \(D\) for every
strict central pair \(p<q\) on strict projection waters and both endpoint
alternatives. The two-endpoint identification argument already handles both
projection waters.

\subsubsection{Equal central peaks}
\label{u26:xii:equal}

Let \(p=q=r\). Projection waters have already been handled. Average one
proper support over both central ties and the low/high ties. It remains
proper with high--low gap \(C>0\) and high--central gap \(A\geq0\), and
globally gives
\begin{equation}
 A\bigl[(x_1-r)+(x_2-r)\bigr]+C(x_3+x_4-2a)\geq0. \label{u26:xii:support}
\end{equation}
The same swap trial lowering central \(1\) by \(\varepsilon\) is feasible
with central peaks \((r-\varepsilon,r)\), lows
\(a+\varepsilon/2\), and highs \(t\). Take the canonical \(U_\varepsilon\)
at these new central peaks and literal \(LLHH\).
\eqref{u26:ENV} fills both centrals and bounds its common low
above by \(a+\varepsilon/2\); the global support
\eqref{u26:xii:support} gives
\begin{equation}
 a+\frac{A\varepsilon}{2C}\leq c_\varepsilon
 \leq a+\frac{\varepsilon}{2}. \label{u26:xii:water}
\end{equation}
For small \(\varepsilon\), its waters have strict projection margins, so
Section~\ref{u26:I-2} also makes \(U_\varepsilon\) a boundary point. It has
distinct central peaks, so the result already proved in
\S\ref{u26:xii:anchors}--\S\ref{u26:xii:two} identifies its whole literal
row. Fix any full \(Q\) at peak \(r\) and use only the first agent's actual
offers \(r-\varepsilon\) at the same five complete opponents
\(Q,L,L,H,H\). Menu closedness and \(\mathsf{OWN}\) pin that first agent at
\(r\) for every full \(P\) at \(r\). Since this holds for every ordered
full pair \((P,Q)\), physical \(\mathsf{SYM}\) applied to the swapped
complete profiles pins the second agent as well. The proper
averaged support gives \(c\geq a\); budget and \(D\)-Pareto completion finish
the entire allocation.

This includes equal zero, positive, and negative peaks.

\subsection{Two interior reports and two reports at each endpoint}\label{u26:XIII}\label{u26:xiii}

For any independent full \(P,Q\), write \(p\leq q\) by a physical
permutation. The goal is
\[
 F(P,Q,L,L,H,H)=U.
\]
The nontrivial
compact case and the notation \([\ell,\bar p]\) are as in Section~\ref{u26:I}.

\subsubsection{The canonical boundary allocation}
\label{u26:xiii:benchmark}

A small vector
\[
 (0,0,-\varepsilon,-\varepsilon,\varepsilon,\varepsilon)
\]
lies in \(X\) and in this peak box. Thus maximal sign volume is positive,
and Section~\ref{u26:I-2} gives the unique vector
\begin{equation}
 U=
 \bigl(\clamp(p;a,t),\clamp(q;a,t),a,a,t,t\bigr),
 \qquad
 \ell\leq a<0<t\leq\bar p. \label{u26:xiii:canonical}
\end{equation}
Zero central peaks are filled. Every central's actual \(L/H\) targets are
the prior \(\mathsf{BT}\)/dual sources, uniformly in the other complete
report. The independent projection closure of \S\ref{u26:iv:projection}
therefore handles \(a=\ell\) or \(t=\bar p\), including all water/peak ties.
Assume henceforth
\[
 \ell<a<0<t<\bar p.
\]
The boundary alternative of Section~\ref{u26:I-2} applies to the two original
literal blocks, irrespective of the central water classification:
\(U\) is a boundary point, for otherwise moving one literal \(L\) and one
literal \(H\) outward by equal amounts would increase feasible peak-box
volume.

\subsubsection{Both central peaks strictly between the waters}
\label{u26:xiii:interior}

If
\[
 a<p\leq q<t,
\]
then \(U=(p,q,a,a,t,t)\). Section~\ref{u26:xii} applies to this boundary
benchmark for the actual complete reports. It covers all normal-gap
alternatives, equal and zero central peaks, and identifies the entire allocation.

\subsubsection{Exactly one saturated central}\label{u26:xiii:one-saturated}

Suppose
\[
 p\leq a<q<t.
\]
At the target obtained by changing \(P\) to literal \(L\), the same feasible
boundary
\[
 U=(a,q,a,a,t,t)
\]
is capped for peaks \((\ell,q,\ell,\ell,\bar p,\bar p)\).
\eqref{u26:CB} identifies its canonical vector directly,
including the original tie \(p=a\). The actual dual \(\mathsf{BT}\) theorem
of Section~\ref{u26:ix} independently identifies that target for every full
\(Q\). Its reporter \(q\) lies in the open dual-\(\mathsf{BT}\) interval
\((A_3,B_3)\): at \(A_3\) the low water would equal \(q\), and at \(B_3\)
the high water would equal \(q\), contrary to \(a<q<t\). Thus
\[
 D_3(q)=a.
\]
The exact \(\mathsf{PLL}\) source of \S\ref{u26:x:promotion} applies to
reporter \(Q\) and retained full \(P,L,L\), with \(p\leq a\), and returns
the whole original row \(U\).

The other exactly-one case
\[
 a<p<t\leq q
\]
follows by reflecting the entire environment, including \(X,F\), and every
complete report. Its canonical rule is the conjugate optimizer.

\par\medskip\noindent\begin{minipage}{\linewidth}
\setlength{\parindent}{2em}
\subsubsection{Both centrals saturated, including all independent shapes and ties}
\label{u26:xiii:both-saturated}

Each central peak is now at most \(a\) or at least \(t\). Replace each
simultaneously by its matching literal endpoint to define an all-endpoint
target with \(l,h\geq2\) and \(l+h=6\). The same
\[
 U=(a^l,t^h),
\]
in that physical endpoint order, is a feasible capped boundary vector for
the target. \eqref{u26:CB} therefore identifies its
canonical value.

Actual \(\mathsf{EB}\) identifies the whole target. Apply the full
mixed-side \(\mathsf{HS}\) theorem of Section~\ref{u26:II-3} to restore the two
independent complete reports \(P,Q\) at their respective thresholds
\(\leq a\) or \(\geq t\). This is its marked-replacement induction: each
replaced agent is pinned by its \(\mathsf{OWN}\) lower-mark endpoint target,
universally over the other complete reports, while the surviving literal
blocks are completed by budget and Pareto efficiency at the current row.
Threshold equality, literal no-ops, and both same-side and opposite-side
restorations are included. Hence the original entire allocation is \(U\).

These cases establish \(S_6(2,2,2)\).

\end{minipage}


\begin{thebibliography}{9}
\providecommand{\natexlab}[1]{#1}
\providecommand{\bibinfo}[2]{#2}
\providecommand{\doi}[1]{\href{https://doi.org/#1}{doi:#1}}
\bibitem[{Barber{\`a} et~al.(2010)Barber{\`a}, Berga and Moreno}]{bbm10}
\bibinfo{author}{Barber{\`a}, S.}, \bibinfo{author}{Berga, D.},
  \bibinfo{author}{Moreno, B.}, \bibinfo{year}{2010}.
\newblock \bibinfo{title}{{Individual versus group strategy-proofness: When do
  they coincide?}}
\newblock \bibinfo{journal}{Journal of Economic Theory} \bibinfo{volume}{145},
  \bibinfo{pages}{1648--1674}.

\bibitem[{Barber{\`a} et~al.(2016)Barber{\`a}, Berga and Moreno}]{bbm16}
\bibinfo{author}{Barber{\`a}, S.}, \bibinfo{author}{Berga, D.},
  \bibinfo{author}{Moreno, B.}, \bibinfo{year}{2016}.
\newblock \bibinfo{title}{{Group strategy-proofness in private good
  economies}}.
\newblock \bibinfo{journal}{American Economic Review} \bibinfo{volume}{106},
  \bibinfo{pages}{1073--1099}.

\bibitem[{Barber{\`a} et~al.(1993)Barber{\`a}, Gul and Stacchetti}]{bgs93}
\bibinfo{author}{Barber{\`a}, S.}, \bibinfo{author}{Gul, F.},
  \bibinfo{author}{Stacchetti, E.}, \bibinfo{year}{1993}.
\newblock \bibinfo{title}{{Generalized median voter schemes and committees}}.
\newblock \bibinfo{journal}{Journal of Economic Theory} \bibinfo{volume}{61},
  \bibinfo{pages}{262--289}.

\bibitem[{Berganti{\~n}os et~al.(2015)Berganti{\~n}os, Mass{\'o} and
  Neme}]{bergantinos15}
\bibinfo{author}{Berganti{\~n}os, G.}, \bibinfo{author}{Mass{\'o}, J.},
  \bibinfo{author}{Neme, A.}, \bibinfo{year}{2015}.
\newblock \bibinfo{title}{{The division problem under constraints}}.
\newblock \bibinfo{journal}{Games and Economic Behavior} \bibinfo{volume}{89},
  \bibinfo{pages}{56--77}.

\bibitem[{Border and Jordan(1983)}]{borderjordan83}
\bibinfo{author}{Border, K.C.}, \bibinfo{author}{Jordan, J.S.},
  \bibinfo{year}{1983}.
\newblock \bibinfo{title}{{Straightforward elections, unanimity and phantom
  voters}}.
\newblock \bibinfo{journal}{Review of Economic Studies} \bibinfo{volume}{50},
  \bibinfo{pages}{153--170}.

\bibitem[{Ching(1994)}]{ching94}
\bibinfo{author}{Ching, S.}, \bibinfo{year}{1994}.
\newblock \bibinfo{title}{{An alternative characterization of the uniform
  rule}}.
\newblock \bibinfo{journal}{Social Choice and Welfare} \bibinfo{volume}{11},
  \bibinfo{pages}{131--136}.

\bibitem[{Moulin(1980)}]{moulin80}
\bibinfo{author}{Moulin, H.}, \bibinfo{year}{1980}.
\newblock \bibinfo{title}{{On strategy-proofness and single peakedness}}.
\newblock \bibinfo{journal}{Public Choice} \bibinfo{volume}{35},
  \bibinfo{pages}{437--455}.

\bibitem[{Moulin(2017)}]{moulin17}
\bibinfo{author}{Moulin, H.}, \bibinfo{year}{2017}.
\newblock \bibinfo{title}{{One-dimensional mechanism design}}.
\newblock \bibinfo{journal}{Theoretical Economics} \bibinfo{volume}{12},
  \bibinfo{pages}{587--619}. \doi{10.3982/TE2307}.

\bibitem[{Sprumont(1991)}]{sprumont91}
\bibinfo{author}{Sprumont, Y.}, \bibinfo{year}{1991}.
\newblock \bibinfo{title}{{The division problem with single-peaked preferences:
  A characterization of the uniform allocation rule}}.
\newblock \bibinfo{journal}{Econometrica} \bibinfo{volume}{59},
  \bibinfo{pages}{509--519}.
\end{thebibliography}
\end{document}